\documentclass[a4paper,11pt]{article}
\pdfoutput=1 % if your are submitting a pdflatex (i.e. if you have
\usepackage{jheppub} % for details on the use of the package, please
\usepackage[T1]{fontenc} % if needed
\usepackage[utf8]{inputenc}
\usepackage{subcaption}
\usepackage{graphicx}
\usepackage{color}

\def\nn{\nonumber}
\def\ov{\overline}

\newcommand{\met}{E_T^{\rm miss}}

\title{\boldmath LHC limits on gluinos and squarks in the minimal Dirac gaugino model}

\author[a]{Guillaume Chalons,}
\author[b]{Mark D. Goodsell,} 
\author[a]{Sabine Kraml,}
\author[a]{Humberto Reyes-Gonz\'{a}lez,}
\author[b]{Sophie L. Williamson}
\affiliation[a]{Laboratoire de Physique Subatomique et de Cosmologie, Universit\'e
  Grenoble-Alpes, CNRS/IN2P3, 53 Avenue des Martyrs, F-38026 Grenoble, France}
\affiliation[b]{Laboratoire de Physique Th\'eorique et Hautes Energies (LPTHE),\\ UMR 7589,
Sorbonne Universit\'e et CNRS, 4 place Jussieu, 75252 Paris Cedex 05, France.}

\emailAdd{chalons@lpsc.in2p3.fr}
\emailAdd{goodsell@lpthe.jussieu.fr}
\emailAdd{sabine.kraml@lpsc.in2p3.fr}
\emailAdd{humberto.reyes-gonzalez@lpsc.in2p3.fr}
\emailAdd{swilliamson@lpthe.jussieu.fr }

\abstract{Dirac gauginos are a well-motivated extension of the MSSM, leading to interesting phenomenological consequences. 
At the LHC, gluino-pair production is enhanced while squark production is suppressed as compared to the MSSM, and 
the decay signatures are altered by a more complex chargino and neutralino spectrum.
We investigate how this impacts current gluino and squark mass limits from Run~2 of the LHC. 
Concretely, we compare different assumptions about the electroweak-ino spectrum through four benchmark models paying particular attention to the effect of the trilinear $\lambda_S$ coupling, which induces a mass splitting between the mostly bino/U(1) adjoint states. Among other results, we show that for large $\lambda_S$ the additional $\tilde\chi^0_2\to f\bar f \tilde\chi^0_1$ decays somewhat weaken the limits on gluinos (squarks) in the case of heavy squarks (gluinos). Moreover, we compare the limits in the gluino vs.\ squark mass plane to those obtained in equivalent MSSM scenarios.}

\notoc
 
\begin{document} 
\maketitle
\flushbottom

\clearpage
%==============================================================================
\section{Introduction}\label{sec:intro}
%==============================================================================

With the current bounds on colourful supersymmetric particles at the LHC, and the consequent implications for naturalness of the Minimal Supersymmetric Standard Model (MSSM), it is timely to consider \emph{non-minimal}  scenarios. A particularly well-motivated extension of the MSSM is to allow Dirac masses for the gauginos, either instead of, or in addition to, Majorana ones. It is the purpose of this paper to derive recent limits on the gluino (fermionic partner of the gluon) and squarks (scalar partners of the quarks) in the minimal Dirac gaugino extension of the MSSM. 

A Dirac term was in fact the original method proposed to allow the gluino to be massive~\cite{Fayet:1978qc}, because the simplest models of global supersymmetry breaking preserve R-symmetry~\cite{Nelson:1993nf} and thus forbid Majorana (but not Dirac) masses; 
this remains an important motivation today. To add Dirac masses for the gauginos, we need to add a Weyl fermion in the adjoint representation of each gauge group; these are embedded in chiral superfields $\mathbf{S}, \mathbf{T}, \mathbf{O}$ which are respectively a singlet, triplet and octet,  and carry zero R-charge. The resulting field content is summarised in Table~\ref{tab:fields}.

The mass terms can then be written by the \emph{supersoft} \cite{Fox:2002bu} operators 
\begin{align}
\mathcal{L}_{\rm supersoft} = \int d^2\theta \Big[\,  \sqrt{2} \, m_{DY} \theta^\alpha \bold{W}_{1\alpha} \bold{S}   
    & + 2 \sqrt{2} \, m_{D2}\theta^\alpha \text{tr} \left( \bold{W}_{2\alpha} \bold{T}\right)  \notag \\
    & \quad +  2 \sqrt{2} \, m_{D3}\theta^\alpha \text{tr} \left( \bold{W}_{3\alpha} \bold{O}\right)\, \Big] + {\rm h.c.}\,,
\label{EQ:supersoft}
\end{align} 
where $\bold{W}_{i\alpha}$ are the supersymmetric gauge field strengths. While it is possible to write the masses through hard breaking operators \cite{Martin:2015eca}, in spontaneously broken SUSY, Dirac masses should only appear through the above \emph{supersoft} terms which have the remarkable property that they do not appear in the renormalisation group (RG) equations for any other operators \cite{Jack:1999fa,Fox:2002bu,Goodsell:2012fm}. %%% \cite{}. 
This means that Dirac gauginos can, in principle, be taken much heavier than their Majorana counterparts since, instead of inducing a logarithmic correction to the sfermion masses, they only induce a finite shift: when this hierarchy is maximally large (i.e.\ we start with zero soft masses for sfermions) it is known as the \emph{supersoft scenario}, which would be realised e.g.\ in models of goldstone gauginos \cite{Alves:2015kia,Alves:2015bba}.

\begin{table}[t]
\begin{center}
{\small 
Chiral and gauge multiplet fields  of the MSSM\\
\begin{tabular}{|c|c|c|c|c|c|}
\hline
  Superfield        & Scalars                  & Fermions & Vectors & ($SU(3)$, $SU(2)$, $U(1)_Y$)  & $R$    \\ \hline
$\mathbf{Q}_i$   & $\tilde{q}_i=(\tilde{u}_{i,L},\tilde{d}_{i,L})$  & $(u_L,d_L)$ & & (\textbf{3}, \textbf{2}, 1/6) & $R_Q$ \\ 
$\mathbf{U}_i$   & $\tilde{u}_{i,R}$              & $u_{i,R}$     & & ($\overline{\textbf{3}}$, \textbf{1}, -2/3) & $2-R_Q - R_H$ \\
$\mathbf{D}_i$   & $\tilde{d}_{i,R}$     & $d_{i,R}$     & & ($\overline{\textbf{3}}$, \textbf{1}, 1/3) & $ R_H - R_Q$ \\ 
$\mathbf{L}_i$    & ($\tilde{\nu}_{i,L}$,$\tilde{e}_{i,L}$) & $(\nu_{i,L},e_{i,L})$ & & (\textbf{1}, \textbf{2}, -1/2) & $ R_L $ \\
$\mathbf{E}_i$   & $\tilde{e}_{i,R}$    & $e_{i,R}$          & & (\textbf{1}, \textbf{1}, 1) & $R_H - R_L$  \\ \hline
$\mathbf{H_u}$  & $(H_u^+ , H_u^0)$ & $(\tilde{H}_u^+ , \tilde{H}_u^0)$ & & (\textbf{1}, \textbf{2}, 1/2) & $R_H$  \\ 
$\mathbf{H_d}$  & $(H_d^0 , H_d^-)$ & $(\tilde{H}_d^0 , \tilde{H}_d^-)$ & & (\textbf{1}, \textbf{2}, -1/2) & $2-R_H$ \\ \hline
$\mathbf{W_{3,\alpha}}$ & & $\lambda_{3} $                       & $G_\mu$              & (\textbf{8}, \textbf{1}, 0) & 1 \\
$\mathbf{W_{2,\alpha}}$ & & $\tilde{W}^0, \tilde{W}^\pm $ & $W^{\pm}_\mu , W^0_\mu$  & (\textbf{1}, \textbf{3}, 0) & 1 \\ 
$\mathbf{W_{Y,\alpha}}$ & & $ \tilde{B} $                       & $B_\mu$              & (\textbf{1}, \textbf{1}, 0 ) & 1  \\ 
\hline
\end{tabular}}\\[4mm]
{\small Additional chiral and gauge multiplet fields in the case of Dirac gauginos}\\
\begin{tabular}{|c|c|c|c|}
\hline
  Superfield               & Scalars, $R=0$                  & Fermions, $R=-1$ & ($SU(3)$, $SU(2)$, $U(1)_Y$) \\ \hline
$\mathbf{O}$ &  $O^a = \frac{1}{\sqrt{2}} (O^a_1 + i O^a_2) $  & $\chi_{O}^a $  & (\textbf{8},\textbf{1},0) \\ 
$\mathbf{T}$ & $T^0 = \frac{1}{\sqrt{2}} (T^0_P + i T^0_M), T^{\pm}$ & $\tilde{W}^{\prime 0}, \tilde{W}^{\prime \pm}$ & (\textbf{1},\textbf{3},0)\\ 
$\mathbf{S}$& $S = \frac{1}{\sqrt{2}} (S_R + i S_I) $ & $ \tilde{B}^{\prime 0} $   & (\textbf{1},\textbf{1},0) \\
\hline
\end{tabular}
\caption{Field content in the Dirac gaugino case.  
Top panel: chiral and gauge multiplet fields of the MSSM; bottom panel: chiral and gauge multiplet 
fields added to those of the MSSM to allow Dirac masses for the gauginos.}
\label{tab:fields}
\end{center}
\end{table}

The supersoft property when applied to the Higgs masses means that Dirac gaugino (DG) models are much more \emph{natural} than Majorana ones, although they do not completely alleviate the little hierarchy problem by themselves \cite{Arvanitaki:2013yja}. On the other hand, the singlet and triplet fields can have new superpotential couplings with the Higgs, 
\begin{align}
W\supset \lambda_S \bold{S} \, \bold{H_u} \cdot \bold{H_d} + 2 \lambda_T \, \bold{H_d} \cdot \bold{T} \bold{H_u} \label{EQ:WNeq2}\,, 
\end{align}
which naturally enhance the Higgs mass at tree level --- and can also be associated with an $N=2$ supersymmetry in the gauge-Higgs sector \cite{Antoniadis:2006uj,Ellis:2016gxa}. An $N=2$ SUSY in turn leads automatically to alignment \cite{Benakli:2018vqz} due to the $SU(2)$ R-symmetry of the two Higgs doublets (which form an $N=2$ hypermultiplet)~\cite{Benakli:2018vjk}. This alignment is surprisingly robust under quantum corrections, where there is an accidental cancellation of $N=2$ breaking effects~\cite{Benakli:2018vqz}.
Moreover, it has been found that the R-symmetry also prevents chirality-flip diagrams, which significantly relaxes flavour constraints \cite{Kribs:2007ac,Fok:2010vk,Dudas:2013gga} and suppresses squark production at the LHC, rendering DG models ``supersafe'' \cite{Heikinheimo:2011fk,Kribs:2012gx,Kribs:2013oda,diCortona:2016fsn}.

The above motivations led to many studies, and realisations being developed \cite{Polchinski:1982an,Hall:1990hq,Fox:2002bu,Nelson:2002ca,Antoniadis:2006uj,%
Amigo:2008rc,Benakli:2008pg,Benakli:2009mk,Benakli:2010gi,%
Carpenter:2010as,Kribs:2010md,Abel:2011dc,Davies:2011mp,Benakli:2011kz,Kalinowski:2011zzc,Frugiuele:2011mh,%
Bertuzzo:2012su,Davies:2012vu,Argurio:2012cd,Argurio:2012bi,Frugiuele:2012pe,%
Frugiuele:2012kp,Benakli:2012cy,Itoyama:2013sn,Chakraborty:2013gea,Csaki:2013fla,Itoyama:2013vxa,Beauchesne:2014pra,%
Bertuzzo:2014bwa,Goodsell:2014dia,Busbridge:2014sha,Chakraborty:2014sda,Diessner:2014ksa,Ding:2015wma,Alves:2015kia,Alves:2015bba,Carpenter:2015mna,Martin:2015eca,Diessner:2015yna,Diessner:2015iln,Diessner:2017ske}. 
The models fall either into the class of those that preserve an exact R-symmetry, or allow a small amount of R-breaking. 
On the former side, the principal example is the Minimal R-Symmetric Supersymmetric Standard Model (MRSSM) \cite{Kribs:2007ac}: this requires the addition of supplementary R-Higgs fields (in the same gauge representation as the MSSM Higgs doublets but with different R-charges) which do not obtain expectation values after electroweak symmetry breaking. However, the couplings in eq.~\eqref{EQ:WNeq2} are forbidden, and the equivalent couplings between the Higgs and R-Higgs fields do not give any tree-level enhancement to the Higgs mass, making the Higgs sector rather like the MSSM --- except that stop mixing is forbidden by the R-symmetry, so that in order to obtain the correct value of the Higgs mass either the new superpotential couplings must be very large \cite{Diessner:2015yna,Diessner:2015iln,Diessner:2017ske} or the stops should be in the $\mathcal{O}(10$--$100)$~TeV range \cite{Benakli:2018vqz}.

Quantum gravity arguments  tell us, however, that no continuous global symmetries should be exact, and so the R-symmetry should be broken at some scale. In this paper, we shall consider the minimal model, often referred to as the Minimal Dirac Gaugino Supersymmetric Standard Model (MDGSSM), described by just the matter content of the MSSM and the adjoint chiral superfields. This model \emph{requires} R-symmetry to be broken in the Higgs sector by a $B_\mu$ term, otherwise it would be spontaneously broken at the same time as electroweak symmetry and generate a massless R-axion in the Higgs sector. As in \cite{Nelson:2002ca,Belanger:2009wf,Benakli:2012cy,Benakli:2014cia,Goodsell:2015ura}, we shall assume that this is the \emph{only} source of R-symmetry breaking, and is motivated by minimality, naturalness (allowing the couplings $\lambda_{S,T}$) and the idea that the Higgs sector couples to a different source of SUSY breaking than the other fields (in order e.g. to generate the $\mu/B_\mu$ terms of similar order etc). This is perfectly consistent at the level of the RG equations: the $B_\mu$ term does not generate other R-breaking operators on RG evolution. This means that the superpotential is
\begin{align}
W^{\rm MDGSSM} =& Y_u^{ij} \,\mathbf{U}_i \mathbf{Q}_j\cdot\mathbf{H_u} 
- Y_d^{ij}\, \mathbf{D}_i \mathbf{Q}_j \cdot\mathbf{H_d} 
- Y_e^{ij}\, \mathbf{E}_i \mathbf{L}_j \cdot\mathbf{H_d}  \nn\\
&  +  \mu\bold{H_u} \cdot \bold{H_d}+ \lambda_S \bold{S} \, \bold{H_u} \cdot \bold{H_d} + 2 \lambda_T \, \bold{H_d} \cdot \bold{T} \bold{H_u} \,, 
\end{align}
where $ \mathbf{Q}_i, \mathbf{L}_j\mathbf{U}_i,\mathbf{D}_i, \mathbf{E}_i, ,\mathbf{H_d} ,\bold{H_u}$ are, respectively, the superfields for the left-handed (LH) squarks; LH sleptons; right-handed (RH) up-type squarks; RH down-type squarks; RH sleptons; down- and up-type Higgs fields as in the MSSM, and  $Y_u^{ij},  Y_d^{ij}, Y_e^{ij} $ which are the standard Yukawa couplings of the MSSM. For the supersymmetry-breaking terms, we add just the supersoft operators eq.~\eqref{EQ:supersoft}, and the \emph{standard} soft terms
\begin{align}
  -\mathcal{L}_{\rm standard\,soft} =  &\hphantom{.} 
  \ov{Q} ^i(m_Q^2)_i^j Q_j + \ov{U} ^i(m_U^2)_i^j U_j + \ov{D} ^i(m_D^2)_i^j D_j + \ov{L} ^i(m_L^2)_i^j L_j + \ov{E} ^i(m_E^2)_i^j E_j \nn\\
  & +m_{H_u}^2 |H_u|^2 + m_{H_d}^2 |H_d|^2 + B_{\mu} (H_u \cdot H_d + \text{h.c.}) \nn \\ 
  & + m_S^2 |S|^2 + 2 m_T^2 \text{tr} (T^{\dagger} T) + m_O^2 |O|^2 \\
  &\hspace{-2.6cm} + \Big[ t_S S+ \frac{1}{2} B_S S^2 + B_T\text{tr}(T T) + B_O\text{tr}(O O) 
    + \frac{A_\kappa}{3}  S^3  + A_{ST}S \mathrm{tr} (TT) +  A_{SO} S \mathrm{tr} (OO) + \text{h.c}\Big]. \nn
\end{align} 

Importantly, the above \emph{contains no SUSY-breaking squark trilinears}; but there is still some small mixing in the stop/sbottom sector due to the $\mu$-term. For simplicity we shall also take $A_\kappa = A_{ST} = A_{SO} = 0$ in the following, which is well justified in gauge mediation models \cite{Benakli:2016ybe}, but we do not expect these parameters to affect our bounds in any significant way. 

Both the MDGSSM and the MRSSM can be embedded in grand unified theories by adding additional electroweak-charged fields~\cite{Benakli:2014cia}; in the former case there is a constrained scenario, the CMDGSSM. For simplicity and generality we shall not include the extra fields, which in any case should not significantly affect the bounds on squarks and gluinos. Instead we shall take a phenomenological approach, choosing masses and couplings at the scale of the colourful superpartners. While the parameter space of such models is large, we shall argue that the constraints we find should be quite general for this class of models. 

The present work will re-examine the LHC bounds on squarks and gluinos in the MDGSSM, which have so far 
been studied only for Run~1 data~\cite{Heikinheimo:2011fk,Kribs:2012gx,Kribs:2013oda}. For the MRSSM there was a study of collider bounds on sleptons and electroweakinos in the MRSSM using Run~1 data \cite{Diessner:2016lvi}, and a recent examination of bounds on charginos in a  gauge-mediation scenario \cite{Alvarado:2018rfl}. 
The scalar octet partners of the gluons, or ``sgluons'', have received more attention in the literature:  
 Dirac gaugino models predict \emph{two} real sgluons, a scalar and pseudoscalar, since they come from a (complex) chiral superfield. These have very interesting collider phenomenology \cite{Plehn:2008ae,Choi:2008ub,Choi:2009ue,Choi:2010gc,GoncalvesNetto:2012nt,Goodsell:2014dia,Chen:2014haa,Beck:2015cga,Kotlarski:2016zhv,Kotlarski:2017dsh}; in particular, if CP is preserved then the pseudoscalar is likely to be relatively light and decay predominantly to tops, so they can be searched for in four-top events \cite{Benakli:2016ybe,Darme:2018dvz}. 

In section \ref{sec:pheno} we give an overview of the phenomenological considerations that shall determine our benchmark scenarios, which we present in section \ref{sec:benchmarks}. %We shall argue that these should be representative of most interesting variations of the MDGSSM, and so the limits obtained should be meaningful. 
We then derive limits on gluino and squark masses first using a simplified models approach in section~\ref{sec:smodels}, before undertaking a full recasting of the fully-hadronic gluino and squark search from ATLAS in section~\ref{sec:recast}. 
A summary and conclusions are given in section~\ref{sec:conclusions}.

%==============================================================================
\section{Phenomenological considerations}\label{sec:pheno}
%==============================================================================

%---------------------------------------------------------------------------
\subsection{Squark and gluino production at the LHC}
%---------------------------------------------------------------------------

\begin{figure}[t]\centering
\includegraphics[width=0.5\textwidth]{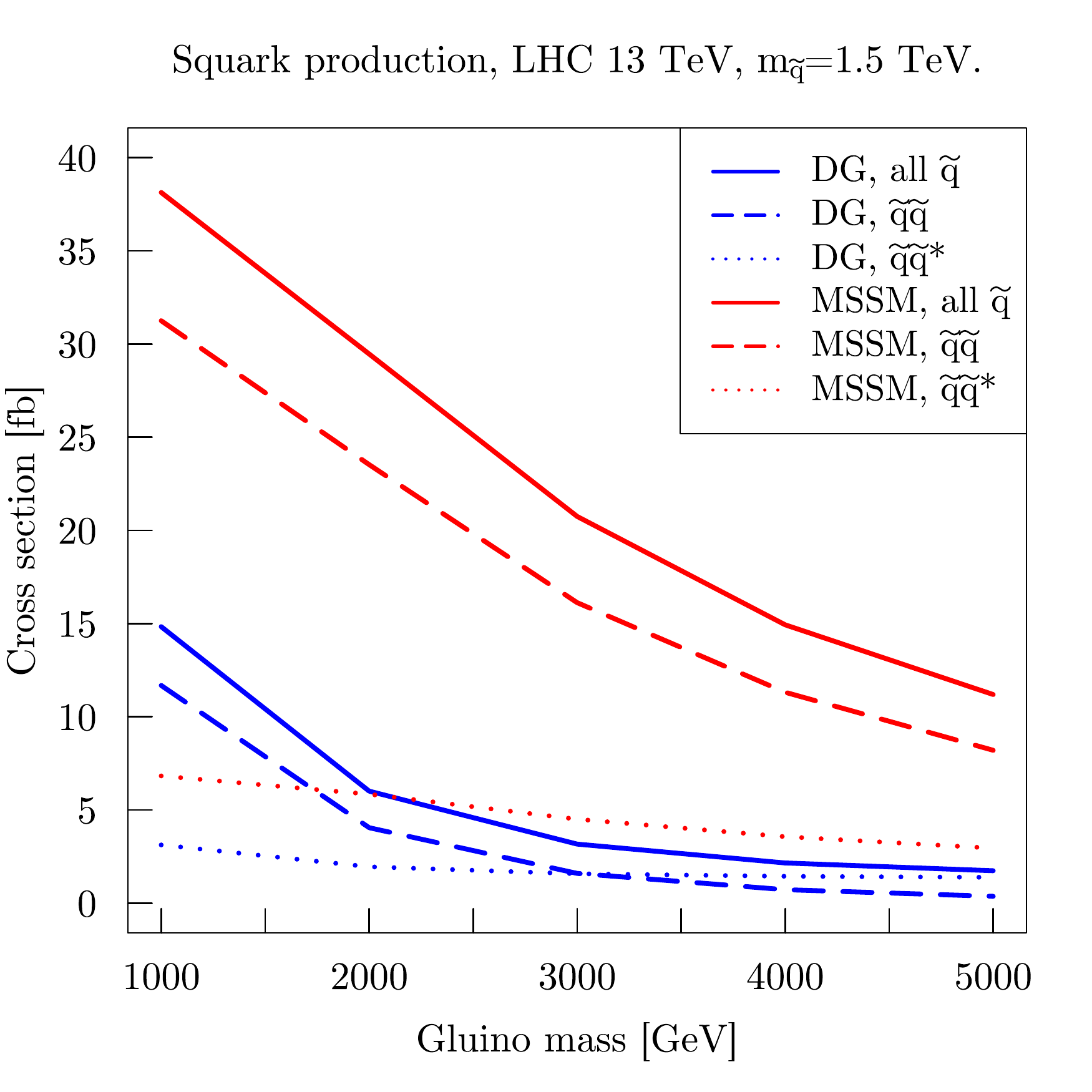}
\caption{\label{fig:squark-production}
Squark production cross-sections at leading order (LO) for the 13 TeV LHC as a function of the gluino mass in the MSSM (in red) and in 
the DG case (in blue), for $m_{\tilde q}=1.5$~TeV, assuming an 8-fold squark degeneracy ($\tilde q=\tilde u,\,\tilde d,\,\tilde c,\,\tilde s$). 
The dashed, dotted and full lines show the squark-squark, squark-antisquark and total squark production cross-sections, respectively.}
\end{figure}

As mentioned above, previous studies of Dirac vs.\ Majorana gauginos highlighted a weakening of collider limits on squarks due to the absence of a chirality flip in the DG case \cite{Heikinheimo:2011fk,Kribs:2012gx,Kribs:2013oda,diCortona:2016fsn}. 
In the MSSM, squark--anti-squark production at the LHC ($pp \to \tilde q_L^{}\tilde q_L^*,\, q_R^{}\tilde q_R^*$) proceeds via $s$-channel gluon 
and $t$-channel gluino exchange; squark--squark production ($pp \to \tilde q\tilde q,\, q^*\tilde q^*$) of same ($LL,\,RR$) 
and mixed ($LR$) chirality via $t$-channel gluino exchange is another important contribution to the total squark production. 
Squark--squark production of same chirality however requires a chirality flip, so it is absent in the DG case. 
Moreover, the other $t$-channel gluino exchange processes are suppressed by $|p|/m_{\tilde g}^2$ in the amplitude, 
where $|p|$ is the momentum in the propagator. This has a huge impact on the total squark production in the 
presence of a heavy Dirac gluino as illustrated in Fig.~\ref{fig:squark-production}.
This suppression of light-flavour  squark production at the LHC is the perhaps best known consequence of Dirac gauginos. 

There are also other interesting consequences, which may impact collider phenomenology. 
For one, the cross-section of gluino-pair production is enhanced in the DG case because of the larger number of degrees of freedom 
than in the MSSM (see \cite{Choi:2008pi} for a detailed discussion).
Another important aspect is the more complex electroweak-ino spectrum.  
Concretely,  while in the MSSM the neutralinos are a linear combination of the four neutral fermions, 
the bino $\tilde{B}$, wino $\tilde{W}^0$ and higgsinos $\tilde{H}^0_u$ and $\tilde{H}^0_d$, in the DG model 
this is supplemented by two adjoint fermions: a bino $\tilde{B}'$ and wino $\tilde{W}'^0$. 
In the chargino sector, the charged winos $\tilde{W}^\pm$ and higgsinos $\tilde{H}^+_u$, $\tilde{H}^-_d$ are 
supplemented by the triplet  $\tilde{W}'^\pm$. 
We thus have six neutralino and three chargino mass eigenstates, which may appear in 
gluino and squark cascade decays. 

One may therefore expect that LHC phenomenology, and constraints from current searches, are different in DG models as
compared to the MSSM.  The purpose of this paper is to investigate what are the concrete LHC limits on gluinos and squarks in the DG case.

%---------------------------------------------------------------------------
\subsection{Electroweak-ino spectrum}\label{sec:EWinos}
%---------------------------------------------------------------------------

The neutralino mass matrix  ${\cal M}_N$ in the basis $(\tilde{B}', \tilde{B}, \tilde{W}'^0,\tilde{W}^0, \tilde{H}_d^0, \tilde{H}_u^0)$ is given by
\begin{align}  
&{\cal M}_N =\nn\\
&\left(\begin{array}{c c c c c c}
0  & m_{DY} & 0     & 0     &  -\frac{ \sqrt{2} \lambda_S }{g_Y}m_Z s_W s_\beta &   - \frac{ \sqrt{2} \lambda_S }{g_Y}m_Z s_W c_\beta  \\
m_{DY} & 0  & 0     & 0     & -m_Z s_W c_\beta &   m_Z s_W s_\beta  \\
0     & 0     & 0  & m_{D2} & - \frac{ \sqrt{2} \lambda_T  }{g_2}m_Z c_W s_\beta & - \frac{ \sqrt{2} \lambda_T  }{g_2}m_Z c_W c_\beta  \\
0     & 0     & m_{D2} & 0   &  m_Z c_W c_\beta & - m_Z c_W s_\beta  \\
-\frac{ \sqrt{2} \lambda_S }{g_Y}m_Z s_W s_\beta & -m_Z s_W c_\beta & -\frac{ \sqrt{2} \lambda_T  }{g_2}m_Z c_W s_\beta &  m_Z c_W c_\beta & 0    & -\mu \\
-\frac{ \sqrt{2} \lambda_S }{g_Y}m_Z s_W c_\beta &  m_Z s_W s_\beta & -\frac{ \sqrt{2} \lambda_T  }{g_2}m_Z c_W c_\beta & -m_Z c_W s_\beta & -\mu & 0    \\
\end{array}\right), 
\label{eq:NeutralinoMassMatrix}
\end{align}
where we denote $s_W=\sin\theta_W$,  $s_\beta=\sin\beta$, $c_\beta=\cos\beta$ and $\tan\beta=v_u/v_d$ the ratio of the Higgs vevs; 
$m_{DY}$ and $m_{D2}$ the bino and wino Dirac masses; $\mu$ the conventional higgsino mass term, and 
$\lambda_S$ and $\lambda_T$ the couplings between the singlet and triplet fermions with the Higgs and higgsino fields. 
The various origins of these mass terms as well as the rotation matrices and eigenvalues are explained in detail in \cite{Belanger:2009wf}. 

Diagonalising eq.~\eqref{eq:NeutralinoMassMatrix}, one ends up with pairs of bino-like, wino-like and higgsino-like neutralinos, 
with small mass splittings {\it within} the bino or wino pairs induced by $\lambda_S$ or $\lambda_T$.\footnote{At least assuming a somewhat hierarchical pattern in $m_{DY}$, $m_{D2}$ and $\mu$; if two or all three mass parameters are close to each other there will be additional effects from sizeable bino, wino and/or higgsino mixing like in the MSSM.} Taking, for instance, $m_{DY}$ sufficiently smaller than $m_{D2}$ and $\mu$, we find a mostly bino/U(1) adjoint lightest SUSY particle (LSP) with a 
mass splitting of 
\begin{equation}
  \Delta m_{\rm LSP} \equiv m_{\tilde\chi^0_2}-m_{\tilde\chi^0_1} 
  = \left| \, 2 \frac{M_Z^2 s_W^2}{\mu} \frac{(2\lambda_S^2 - g_Y^2 )}{g_Y^2}c_\beta s_\beta \, \right|  \,.
\end{equation}
For the models that we shall consider,  this can go up to tens of GeV. 

Turning to the charged fermions, there are three charginos $\tilde\chi^\pm_{1...3}$ from a linear 
combination of the charged higgsinos, $\tilde{H}^+_u$, $\tilde{H}^-_d$, charged gauginos $\tilde{W}^\pm$ and adjoint $\tilde{W}'^\pm$. 
In the basis $v^+ = (\tilde{W}'^+,\tilde{W}^+,\tilde{H}^+_u)$,  $v^- = (\tilde{W}'^-,\tilde{W}^-,\tilde{H}^-_d)$, 
the chargino mass matrix is 
\begin{equation}
{\cal M}_C = 
\left(\begin{array}{c c c}
 0  & m_{2D} &\frac{ {2} \lambda_T  }{g} m_W c_\beta \\
m_{2D} & 0   & \sqrt{2} m_W s_\beta \\
- \frac{ {2} \lambda_T  }{g} m_W s_\beta & \sqrt{2} m_W c_\beta & \mu \\
\end{array}\right) \,, 
\label{diracgauginos_CharginoMassarray}
\end{equation}
where we again assumed that Majorana mass terms are absent. 
This gives one higgsino-like $\tilde\chi^\pm$ and two wino-like $\tilde\chi^\pm$ -- the latter ones again with a small splitting driven by $\lambda_T$.

The possible impact on collider phenomenology becomes apparent when considering that gluino and squark decays will be shared out 
over the different neutralino and chargino states with small mass splittings. 
For instance, for a mostly bino/U(1) adjoint LSP, $\tilde q_R^{}\to q\tilde\chi^0_1$ or $q\tilde\chi^0_2$ with roughly equal   
branching ratios. 
If $\Delta m_{\rm LSP}<m_Z$, the $\tilde\chi^0_2$ then decays to $f\bar f \,\tilde\chi^0_1$ via an off-shell $Z$-boson.  
Therefore, while in the MSSM with a bino-like LSP $pp\to \tilde q_R^{}\tilde q_R^{}$ leads to events with 2~jets + $\met$, 
in the DG model with somewhat split binos, we may get a mix of events with 2, 4 or 6 jets + $\met$, and with a small rate 
also jets + $\ell^+\ell^-$ + $\met$. We note also that, due to $Z^*\to \nu\bar \nu$, some of the $\tilde\chi^0_2$ decays will be invisible.
%Likewise, the $\tilde q_L^{}$ will share out its decays over the wino-like neutralino and chargino states, which will then 
%cascade-decay to the $\tilde\chi^0_{1,2}$, leading to events with jets + $W$, $Z$ or Higgs bosons + $\met$. 
Similar considerations apply to all SUSY cascade decays. % and to the third generation (stops and sbottoms). 

Finally, %it is important to note that 
the mass splitting between the two lightest neutralinos determines the $\tilde\chi^0_2$ lifetime. 
If the splitting is very small, the $\tilde\chi^0_2$ can live long enough to effectively be a co-LSP on collider scales and appear only  as $\met$.
For larger mass splittings, the $\tilde\chi^0_2$ can decay promptly, leading to the complex signatures discussed in the paragraphs above.
In between, the $\tilde\chi^0_2$ is a long-lived neutral particle, whose decays can give signatures with displaced vertices.

%---------------------------------------------------------------------------
\subsection{Effect of R-symmetry breaking}
%---------------------------------------------------------------------------

The mass-splittings in the neutralinos are due to the R-symmetry breaking effect of both the $H_u$ and $H_d$ fields obtaining an expectation value -- hence they are proportional to $c_\beta s_\beta$ which vanishes for large and small $\tan \beta$. In addition, when $\lambda_S = g_Y/\sqrt{2}, \lambda_T = g_2/\sqrt{2}$, we have an effective global symmetry among the gauginos and higgsinos which allows the neutralinos and charginos to remain of Dirac type at tree-level -- this is not actually 
the $SU(2)$ R-symmetry, of which the higgsinos are actually singlets.  

This means that any Majorana masses for the neutralinos and charginos (which we are neglecting) should be smaller than the above splittings in order for the analysis in this paper to be valid: this makes a difference to the softness of the decays from $\tilde{\chi}_2^0$ to $\tilde{\chi}_1^0$, for example. 

%At tree-level the gluino states in our model are exactly Dirac, 
Turning to the gluinos, at tree level $\tilde g_{1,2}$ are exactly Dirac in our model; 
the two states are only split by a tiny difference at one loop from the small amount of mixing between the left- and right-handed squarks proportional to $\mu$. Here, however, a modest Majorana mass could be tolerated, since the only effect would be to split the eigenstates and so be distinguishable in a detector as separate particles: in our benchmarks they shall be indistinguishable. Interestingly, in our model the octet fermion $\chi_O$ only couples to the scalar octets, gluino and gluons. 
Hence the two gluino mass eigenstates, $\tilde{g}_1, \tilde{g}_2 = \frac{1}{\sqrt{2}} (\lambda_3 + \chi_O), \frac{i}{\sqrt{2}} (\lambda_3 - \chi_O) $, couple only to the squarks and quarks through the component $\lambda_3$, and their couplings are the same up to a factor of $i$. This means that over the parameter space, their decays are \emph{almost} identical, meaning that together they behave like a purely Dirac gluino---except for when the decay is highly non-relativistic.

In our model, the only relevant non-relativistic two-body decays of a gluino are when a squark becomes nearly degenerate with it; and so to obtain differences between $\tilde{g}_1$ and $\tilde{g}_2$ decays we would furthermore need a sizeable source of R-symmetry breaking, which means squark mixing. %-- and thus decays involving stops/sbottoms just below the gluino mass. 
We can therefore expect a sizeable difference between the two gluino decays into stops or sbottoms only near the kinematic limit.  This can be seen as follows: 
for a two-body decay $\tilde{g}_i \rightarrow q \tilde{q}$ for $i=1,2$ we can write the couplings (suppressing the gauge and Lorentz indices) as 
\begin{align}
\mathcal{L} \supset& - \sqrt{2} g_3 \tilde{q}^*_L q_L^{} \lambda_3 +  \sqrt{2} g_3  \ov{q}_R^{} \tilde{q}_R^{} \ov{\lambda}_3 
\end{align}
and so if $\tilde{q}_L^{} = \cos \theta_q\tilde{q}_1^{} + \sin \theta_q\tilde{q}_2^{}$,  
$\tilde{q}_R^* = -\sin \theta_q\tilde{q}_1^{} + \cos \theta_q\tilde{q}_2^{}$, then the coupling to say $\tilde{q}_1^{}$ is 
\begin{align}
  \mathcal{L} \supset& - \tilde{q}_1^* \Big[ c_L^i (q \tilde{g}_i) + c_R^i (\ov{q} \ov{\tilde{g}}_i) \Big], \qquad 
  c_L^1 = \sqrt{2} g_3 \cos \theta_q, \quad c_R^1 = -\sqrt{2} g_3 \sin \theta_q,
\end{align}
while  $c_L^2 = -i c_L^1, (c_R^2)^* = -i c_R^1$.  
The width for the gluino decays is then
\begin{align}
  \Gamma ( \tilde{g}_i \rightarrow q \tilde{q}_i) =& \frac{K}{32\pi m_{\tilde{g}_i}^3} \bigg[(m_{\tilde{g}_i}^2 + m_q^2 - m_{\tilde{q}_i}^2)  ( |c_L|^2 + |c_R|^2) + 2 m_q m_{\tilde{q}_i} (c_L^* c_R + c_R^* c_L) \bigg], \nn\\
K \equiv& \sqrt{(m_{\tilde{g}}^2 - m_q^2 - m_{\tilde{q}_i}^2)^2 - 4 m_q^2 m_{\tilde{q}_i}^2}.
\end{align}
So then when $m_{\tilde{g}_i} \sim m_q + m_{\tilde{q}_i}, m_{\tilde{q}_i} \gg m_q$, we have $(m_{\tilde{g}_i}^2 + m_q^2 - m_{\tilde{q}_i}^2 )\simeq 2 m_q m_{\tilde{q}_i} $ and
\begin{align}
  \Gamma ( \tilde{g}_i \rightarrow q \tilde{q}_i) \simeq&  \frac{K m_q g_3^2}{16\pi m_{\tilde{g}_i}^2} \bigg[ 1 \pm 2\cos \theta_q \sin \theta_q \bigg].
\end{align}
Hence for maximal squark (stop or sbottom) mixing there is a complete suppression of one of the decays in this limit. 

For three-body decays of a gluino to neutralinos and quarks, we shall argue below that in our model the neutralinos should be light, and so even though the neutralinos themselves significantly break the R-symmetry through their mixings, the quarks/neutralinos should be relativistic and we should not see a significant difference between the two gluino components.

%---------------------------------------------------------------------------
\subsection{Model constraints}
%---------------------------------------------------------------------------

As mentioned above, the limits on gluino and (first/second generation) squark masses %must 
depend on the other parameters in the model, in particular the mass of the lightest supersymmetric partner, but also on the details of the decay chains. In the (phenomenological) MSSM, it is reasonable to consider the bino/wino/higgsino masses as free parameters. However, in the MDGSSM (and in DG models generally) these have a large effect on the Higgs mass at tree level. Indeed, it is well known that in the supersoft limit the Higgs D-term potential is erased \cite{Fox:2002bu}; and a large $\mu$-term has a similar effect. Moreover, the singlet and triplet scalars obtain tree-level masses $m_{SR}, m_{TP}$ proportional to the Dirac mass terms:
\begin{align}
  m_{SR}^2 = m_S^2 + 4 |m_{DY}|^2 + B_S, \qquad m_{TP}^2 = m_T^2 + 4 |m_{D2}|^2 + B_T,  
\end{align}
and so if $m_{DY}$ or $m_{D2}$ are large then the scalar singlet/triplet should be heavy. If we then integrate them out, then the correction to the Higgs quartic coupling is
\begin{align}
  \delta \lambda \sim \mathcal{O} \left(\frac{g_Y m_{DY}}{m_{SR}}\right)^2 
  + \mathcal{O} \left(\frac{\sqrt{2}\lambda_S  m_{DY}}{m_{SR}}\right)^2 
  + \mathcal{O} \left(\frac{g_2 m_{D2}}{m_{TP}}\right)^2 
  + \mathcal{O} \left(\frac{\sqrt{2}\lambda_T  m_{D2}}{m_{TP}}\right)^2,
\end{align}
The exact expressions for the Two-Higgs Doublet model parameters are given in \cite{Benakli:2018vqz}. This means that we need to make the singlet and triplet scalars heavy \emph{relative to the gauginos and higgsinos} in order to not suppress the Higgs mass or even render the potential unstable. Without removing the scalars from the spectrum entirely and losing all trace of naturalness,  this means keeping the gauginos/higgsinos well below a TeV. 

Additionally, scalar triplet fields are well-known to generate a shift to the electroweak $\rho$-parameter at tree-level:
\begin{align}
   \Delta \rho = \frac{\Delta m_W^2}{m_W^2} = \frac{v^2}{m_{TP}^4} \bigg( \sqrt{2} \lambda_T \mu + g_2 m_{D2} c_{2\beta}  \bigg)^2 ,
\end{align}
while the experimental best-fit value is \cite{PDG}
\begin{align}
  \Delta \rho = (3.7 \pm  2.3) \times 10^{-4},
\end{align}
leading to $m_{TP} \gtrsim 2$~TeV for typical values of $\mu, m_{D2} \sim 500$~GeV. Numerically we find it is hard to find satisfactory parameter points for gaugino/higgsino masses of $\mathcal{O}({\rm TeV})$ and so in our benchmark points we shall take them to be only a few hundred GeV. 

On the other hand, in the decoupling limit, the light Higgs mass is given by
\begin{align}
m_{h_1}^2 \simeq M_Z^2 c_{2\beta}^2 + \frac{(\lambda_S^2 + \lambda_T^2)}{2} v^2 s_{2\beta}^2 + ...  
\end{align}
and so taking small $\tan \beta$ and moderate values of $\lambda_S, \lambda_T$ we can enhance the Higgs mass at tree-level without having exceptionally heavy stops (given that the stop mixing will be small in the absence of SUSY-breaking trilinear couplings).

%==============================================================================
\section{Benchmark scenarios}\label{sec:benchmarks}
%==============================================================================

We have argued that a typical MDGSSM scenario should have electroweakinos of $\mathcal{O}(500)$~GeV, a triplet scalar heavier than $2$~TeV, and if we want to enhance naturalness of the model (avoiding stop masses larger than $\mathcal{O}(10)$~TeV), small $\tan \beta$ and $2(\lambda_S^2 + \lambda_T^2) > g_Y^2 + g_2^2$. For the sake of simplicity, and since we have no reason to suspect a large splitting 
%of the first two generations of squark masses, 
of left- and right-chiral squarks, 
we shall take $m_{Q_i}^2 = m_{U_i}^2 = m_{D_i}^2 $, and take a common value for the first two generations, while allowing the third generation squark masses to vary so as to obtain the correct Higgs mass (some stop contribution is necessary unless we take values of $\lambda_S, \lambda_T$ that are large).

\begin{figure}[t!]\centering
\includegraphics[width=0.5\textwidth]{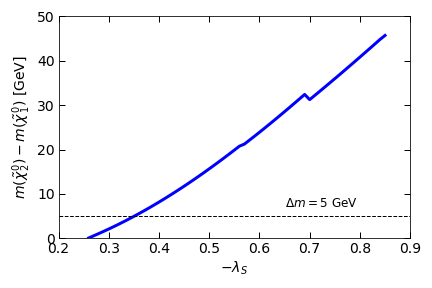}\includegraphics[width=0.5\textwidth]{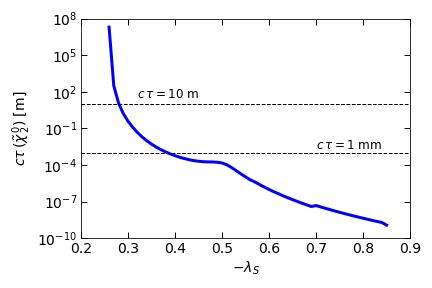}
\caption{Influence of $\lambda_S$ on the mass splitting between the two bino-like mass eigenstates $\tilde\chi^0_{1,2}$ (left) and 
on the lifetime of the $\tilde\chi^0_{2}$ (right) for the parameters of eq.~\eqref{eq:EWinoparameters} and $\tan\beta=2$.}
\label{fig:lamSdependence}
\end{figure}

To quantitatively investigate how this influences the LHC limits, we choose four benchmark scenarios, 
with different values of $\lambda_S$, $\lambda_T$. 
Concretely  we take $m_{DY}<\mu<m_{2D}$ with, for the first three benchmarks,
\begin{equation}
    m_{DY}=200~{\rm GeV},\quad \mu=400~{\rm GeV},\quad m_{D2}=500~{\rm GeV}.
    \label{eq:EWinoparameters}
\end{equation} 
Moreover, to favor a large tree-level boost to $m_{h_1}$, we take $\tan\beta=2$. %; then stop masses of the order of 1.5~TeV are sufficient to achieve $m_{h_1}=125$~GeV. 
This gives a hierarchical spectrum of bino-, higgsino- and wino-like states with masses of about 200, 400 and 500~GeV, respectively. 
Finally, we set $\lambda_T=0.2$ and choose two values of $\lambda_S$, $\lambda_S=-0.27$ and $-0.74$, 
to have cases with small and sizeable $\tilde\chi^0_{1,2}$ mass splittings. 
The dependence of the $\tilde\chi^0_{1,2}$ mass splitting and the $\tilde\chi^0_{2}$ lifetime on $\lambda_S$
is shown in Fig.~\ref{fig:lamSdependence}. 

With this setup, the masses of gluinos and squarks are treated as free parameters ($m_{3D}$ and a scalar soft mass-squared parameter), 
while the masses of the 3rd generation squarks are adjusted such that $m_{h_1}\in [123,\,127]$~GeV. 
The calculation of the mass spectrum and decay branching ratios is done with {\tt SARAH}~\cite{Staub:2012pb,Staub:2009bi,Staub:2010jh,Staub:2012pb,Staub:2013tta,Goodsell:2017pdq}  
and {\tt SPheno}~\cite{Porod:2011nf}, including Higgs mass calculation at the 2-loop level~\cite{Goodsell:2014bna,Goodsell:2015ira,Braathen:2017izn}.  
We consider three distinct cases:\footnote{We note that we do not consider any dark matter constraints here. This is justified as we are interested in unequivocal collider constraints on the colored sector without assumptions on the cosmological history of the universe. For a discussion of DG dark matter within standard cosmology, see \cite{Belanger:2009wf}.}
\begin{align}
  {\rm DG1:}\quad &  \lambda_S = -0.27; \; m_{\tilde{t}} \sim m_{\tilde{b}} \sim 3.6 \, \text{TeV}, \label{eq:BM1} \\
  {\rm DG2:} \quad & \lambda_S = -0.74; \; m_{\tilde{t}} \sim m_{\tilde{b}} \sim 2.6 \, \text{TeV},  \label{eq:BM2} \\
  {\rm DG3:} \quad & \lambda_S = -0.74; \; m_{\tilde{t}} \sim m_{\tilde{b}} \sim 1.6 \, \text{TeV}.  \label{eq:BM3}
\end{align}
For DG1 with $\lambda_S=-0.27$, the two bino-like mass eigenstates $\tilde\chi^0_{1,2}$ are quasi-degenerate with sub-GeV mass splitting, 
and the $\tilde\chi^0_{2}$ has a mean decay length of nearly 3~km, 
so that it will appear as a co-LSP. 
For $\lambda_S=-0.74$ (DG2 and DG3), the two bino-like mass eigenstates $\tilde\chi^0_{1,2}$ have masses of about 
182~GeV and 216--218~GeV, respectively, and the $\tilde\chi^0_{2}$ decays promptly into $\tilde\chi^0_{1}\,f\bar f$ via an off-shell $Z$.

\begin{table}[t]\centering
\begin{tabular}{|c|c|c|c|c|}
\hline
\multicolumn{5}{|c|}{Parameters}\\
\hline
 & DG1 & DG2 & DG3 & DG4 \\
\hline
$m_{1D}$  & 200 & 200 & 200 & 200  \\
$m_{2D}$  & 500 & 500 & 500 & 1175  \\
$\mu$  & 400 & 400 & 400 & 400  \\
$\tan\beta$  & 2 & 2 & 2 & 2  \\
$-\lambda_S$ & 0.27 & 0.74 & 0.74 & 0.79  \\
$\sqrt{2}\,\lambda_T$ & 0.14 & 0.14 & 0.14 & $-0.26$  \\
$m^2_{\tilde Q_3}$  & 1.25e7 & 6.5e6 &  2.26e6 &  8.26e6  \\
\hline
$m^2_{\tilde Q_1}$  & 6.25e6 & 6.25e6 &  6.25e6 &  6.25e6  \\
$m_{3D}$  & 1750 & 1750 & 1750 & 1750  \\
\hline
\end{tabular}
\begin{tabular}{|c|c|c|c|c|}
\hline
\multicolumn{5}{|c|}{Masses}\\
\hline
 & DG1 & DG2 & DG3 & DG4 \\
\hline
$\tilde\chi^0_1$ & 201.35 & 182.1 & 181.8 & 182.4  \\
$\tilde\chi^0_2$ & 201.72 &  218.0 & 216.6 & 213.2  \\
$\tilde\chi^0_3$ & 403 & 400 & 396 & 408  \\
$\tilde\chi^0_4$ & 419 & 445 & 441 & 437  \\
$\tilde\chi^0_5$ & 537 & 536 & 535 & 1226  \\
$\tilde\chi^0_6$ & 548 & 548 & 546 & 1227  \\
$\tilde\chi^\pm_1$ & 400 & 395 & 391 & 398  \\
$\tilde\chi^\pm_2$ & 536 & 536 & 534 & 1224  \\
$\tilde\chi^\pm_3$ & 549 & 548 & 547 & 1229  \\
$\tilde t_1$ & 3604 & 2607 & 1590 & 2894  \\
$\tilde t_2$ & 3613 & 2637 & 1613 & 2927  \\
$h_1$ & 124.0 & 125.0 & 125.3 & 125.2  \\
\hline
\end{tabular}
\caption{Parameters and masses (in GeV) of the four benchmark scenarios; $m_{1D}$, $m_{2D}$, $\mu$, $\tan\beta$,  $\lambda_S$, $\lambda_T$
and the soft masses of the third generation ($m^2_{\tilde Q_3}=m^2_{\tilde U_3}=m^2_{\tilde D_3}$) are fixed for each benchmark, 
while $m_{3D}$ and $m^2_{\tilde Q_1}=m^2_{\tilde U_1}=m^2_{\tilde Q_1}$ will be varied to scan over gluino and squark masses. 
The sgluons have masses of about 1.6 and 3.9~TeV and play no role for the phenomenology discussed here.}
\label{tab:benchmarks}
\end{table}

\begin{table}[t]\centering
\begin{tabular}{|l|c|c|c|c||c|c|}
\hline
 & DG1 & DG2 & DG3 & DG4 & MSSM1 & MSSM4\\
\hline\hline
\multicolumn{7}{|c|}{Gluino decays, $m_{\tilde g}\approx 2$~TeV, $m_{\tilde q}\approx 2.6$~TeV}\\
\hline
$\tilde g\to q\bar q$ + binos & 12\%  & 6\% &  --  & 18\%  & 10\% & 15\% \\
$\tilde g\to b\bar b$ + binos & --  & 1\% &  --  & 6\% & -- &  1\%  \\
$\tilde g\to t\bar t$ + binos & 1\%  & 4\% &  -- & 6\% & 1\% &  3\%  \\
$\tilde g\to (q\bar q^{(')},\,b\bar b)$ + heavy EW-inos & 66\%  & 36\% &  -- & 13\% & 66\%  &  19\% \\
$\tilde g\to (t\bar t,\, t\bar b, b\bar t)$ + heavy EW-inos & 20\%  & 53\% &  -- & 61\% &   23\% &  62\% \\
$\tilde g\to t+\tilde t_{1,2}$  &  -- & -- &  48\%  &   -- & -- & --\\
$\tilde g\to b+\tilde b_{1,2}$  & --  & --  & 52\%  & -- &   -- & --\\
\hline\hline
\multicolumn{7}{|c|}{Squark decays, $m_{\tilde q}\approx 2$~TeV, $m_{\tilde g}\approx 2.6$~TeV}\\
\hline
$\tilde q_R \to q$ + binos & 99\%  & 99\%   & 98\% & 99\% & 92\% & 92\%  \\
%$\tilde q_L \to q$ + binos &   &   &   & &  2\% & 3\% \\
$\tilde q_L \to q$ + heavy EW-inos & 99\%  & 99\%  & 99\%  & 97\% &  98\% & 97\% \\
\hline
\end{tabular}
\caption{Branching ratios of gluino and squark decays for DG1--DG4. 
For gluino decays we consider the mass hierarchy $m_{\tilde q}< m_{\tilde g_{1,2}}$, 
for squark decays  the mass hierarchy $m_{\tilde \tilde g_{1,2}} > m_{\tilde q}$. 
The columns MSSM1 and MSSM4 give the comparison to the equivalent MSSM case 
with $M_1=200$~GeV, $\mu=400$~GeV and $M_2=500$~GeV (MSSM1) or 1200~GeV (MSSM4); 
third generation squark masses are about 3.6~TeV for MSSM1 and 3~TeV for MSSM4, while $\tan\beta=10$ 
and $A_t=-4$~TeV to achieve $m_{h_1}\approx 125$~GeV. }
\label{tab:branchings}
\end{table}

Since we are mostly interested in gluino and squark cascade decays, we consider also a fourth benchmark with heavy winos by moving $m_{2D}$ above 1~TeV, thus on the one hand somewhat suppressing decays into wino-like states, and on the other hand changing the kinematic distributions of such cascades. Concretely,  
\begin{align}
  {\rm DG4:}\quad &   m_{1D}=200~{\rm GeV},\; \mu=400~{\rm GeV},\; m_{2D}=1175~{\rm GeV},\; \notag\\ 
   & \lambda_S = -0.79,\; \lambda_T = -0.37,\; m_{\tilde{t}} \sim m_{\tilde{b}} \sim 3 \, \text{TeV}. \label{eq:BM4} 
\end{align}
The main parameters and resulting masses for the four benchmark scenarios are summarised in Table~\ref{tab:benchmarks}. 
Examples of gluino and squark decay branching ratios are given in Table~\ref{tab:branchings} and compared to the 
branching ratios in the MSSM with an equivalent bino/wino/higgsino spectrum. 

The complete SLHA spectrum files produced with {\tt SARAH/SPheno} are available at \cite{zenodo:bm-dataset}.\footnote{For the sake of reproducibility of our results, we provide moreover the {\tt SPheno} model and input files, as well as the UFO model and two helpful scripts for modifying the {\tt SPheno} .spc files so they can be used for event generation with {\tt MadGraph}/{\tt Pythia}.} 
Note here, that our conventions differ (as usual) from the {\tt SARAH\ DiracGauginos} implementation. We have
\begin{equation}
\begin{array}{|c|c|} \hline
\mathrm{Parameter} & {\tt SARAH}\ \mathrm{convention}  \\\hline
\lambda_S & -{ \tt lam} \\
\lambda_T &  {\tt LT}/\sqrt{2} \\\hline\end{array}
\end{equation}

\bigskip

\begin{figure}[t]\centering
\includegraphics[width=0.76\textwidth]{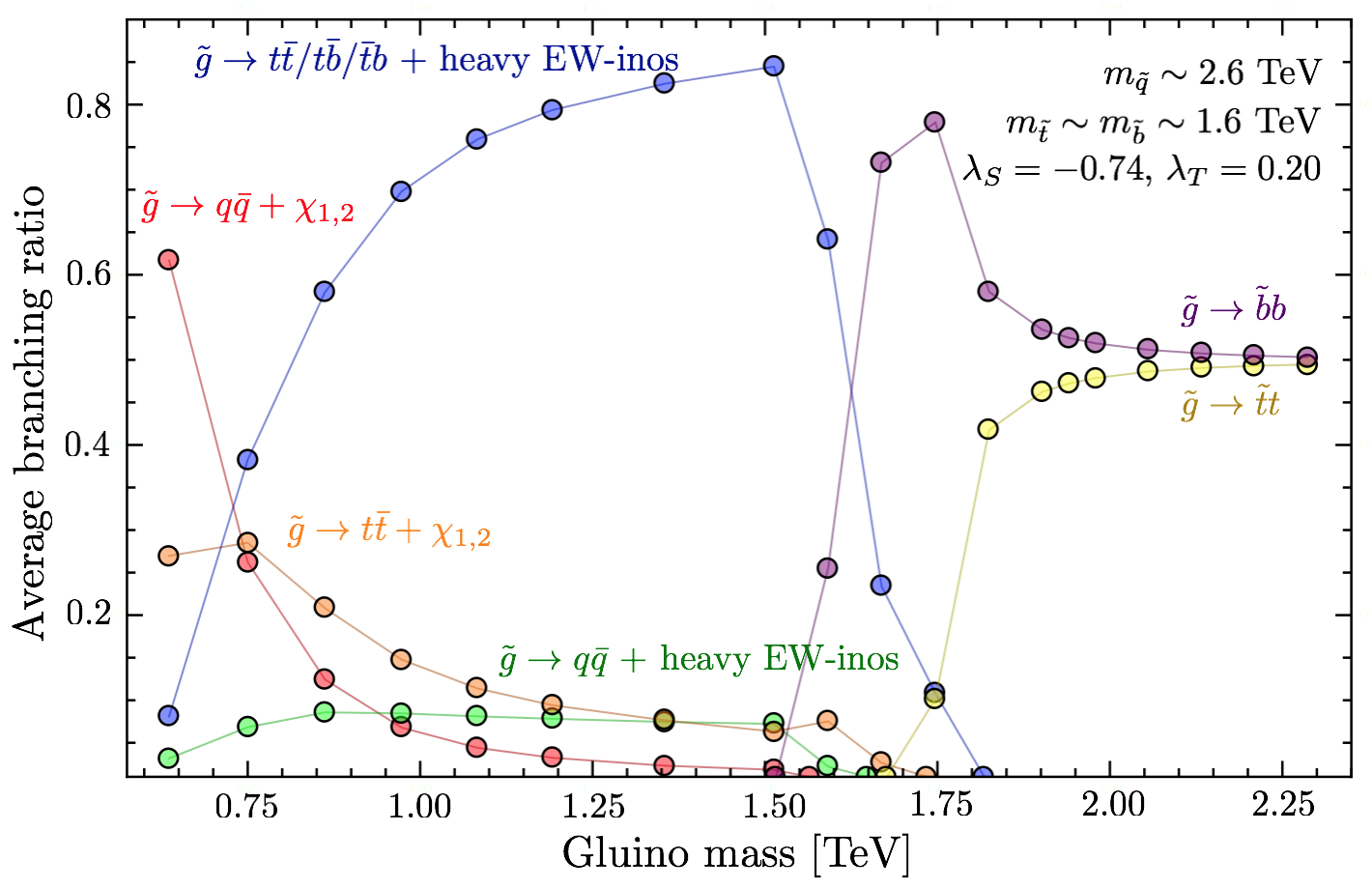}
\caption{Branching ratios of gluino decays (averaged over $\tilde g_1$ and $\tilde g_2$) for DG3 as function of the gluino mass, 
for $m_{\tilde q}\approx 2.6$~TeV. }
\label{fig:branchings-gluino-ext}
\end{figure}

Scenarios DG1, DG2 and DG3 have heavy stops and sbottoms, so the gluino branching ratios in Table~\ref{tab:branchings} 
will not change significantly with the gluino mass in the region accessible with current LHC data, as long as $m_{\tilde g}<m_{\tilde q}$
(if $m_{\tilde g}>m_{\tilde q}$, then of course $\tilde g\to q\tilde q$ decays dominate). This is different for DG3 which has stops and sbottoms 
at about 1.6~TeV. Here the gluino branching ratios vary a lot with  $m_{\tilde g}$ up to 2~TeV, 
as shown in Fig.~\ref{fig:branchings-gluino-ext}. 
We note that in this figure BR($\tilde g_1$) and BR($\tilde g_2$) are averaged over  because R-symmetry breaking 
effects lead to differences in $\tilde g_1$ and $\tilde g_2$ decays near the threshold where 2-body decays into sbottoms/stops  
become kinematically allowed. These differences are however experimentally not observable. 

%==============================================================================
\section{Simplified model limits}\label{sec:smodels}
%==============================================================================

Within the MSSM, ATLAS and CMS have excluded gluino (light-flavor squark) masses up to about 1800--2025 (1550) GeV 
assuming decoupled squarks (gluinos) and a single decay channel into the neutralino LSP with 100\% branching ratio~\cite{Aaboud:2017vwy,Sirunyan:2017kqq}. 
In the DG case, the twice as large gluino production cross-section should increase the gluino mass limit by about 150--200~GeV;
the bound on squark masses remains the same, since the quoted MSSM limit is already for decoupled gluinos.  

The constraints which can be derived in the context of such ``simplified models'' considerably weaken 
in realistic scenarios where the gluinos (squarks) share out their branching ratios over several decay channels~\cite{Ambrogi:2017lov}.\footnote{This is in particular the case if only cross-section upper limits are available for simplified model spectra. Efficiency maps for all signal regions for a large enough set of simplified models would allow us to combine the contributions from different signal topologies in the simplified model approach \cite{Ambrogi:2017neo}.}  
For instance, if ${\rm BR}(\tilde g \to q\bar q \tilde\chi^0_1)=0.1$, only 1\% of the total gluino-pair production is constrained 
by the $pp\to \tilde g\tilde g$, $\tilde g \to q\bar q \tilde\chi^0_1$ simplified model %(in the following denoted as T1) 
upper limits. 
Likewise, if $\tilde q_L^{}$ decay via heavy EW-inos, only $\tilde q_R^*\tilde q_R^{}$ production is effectively constrained by the $pp\to \bar{\tilde q}\tilde q$, $\tilde q \to q\tilde\chi^0_1$ simplified model limits. 
On the other hand, the production cross-sections themselves can be [much] larger than in the simplified model picture, if 
gluino (squark) contributions to squark (gluino) production are not decoupled in the parameter space we are interested in.

To illustrate explicitly the consequences for our benchmark scenarios, we scan over gluino and squark masses for two cases, 
DG1 and DG3, and evaluate the simplified model constraints with 
{\tt SModelS}~\cite{Kraml:2013mwa,Ambrogi:2017neo}. Here we use the v1.1.2 database of {\tt SModelS}, 
which includes the Run~2 SUSY search results for 36~fb$^{-1}$ from CMS as detailed in \cite{Dutta:2018ioj}. 
The decay branching ratios are again computed with {\tt SARAH}/{\tt SPheno}. 
Cross-sections are computed at leading order with {\tt MadGraph5\_aMC@NLO}~\cite{Alwall:2014hca} 
using the Dirac gaugino UFO model of~\cite{Staub:2012pb}. 
(The effect of higher-order corrections will be commented on in the next section.)

\begin{figure}[t!]\centering
\includegraphics[width=0.5\textwidth]{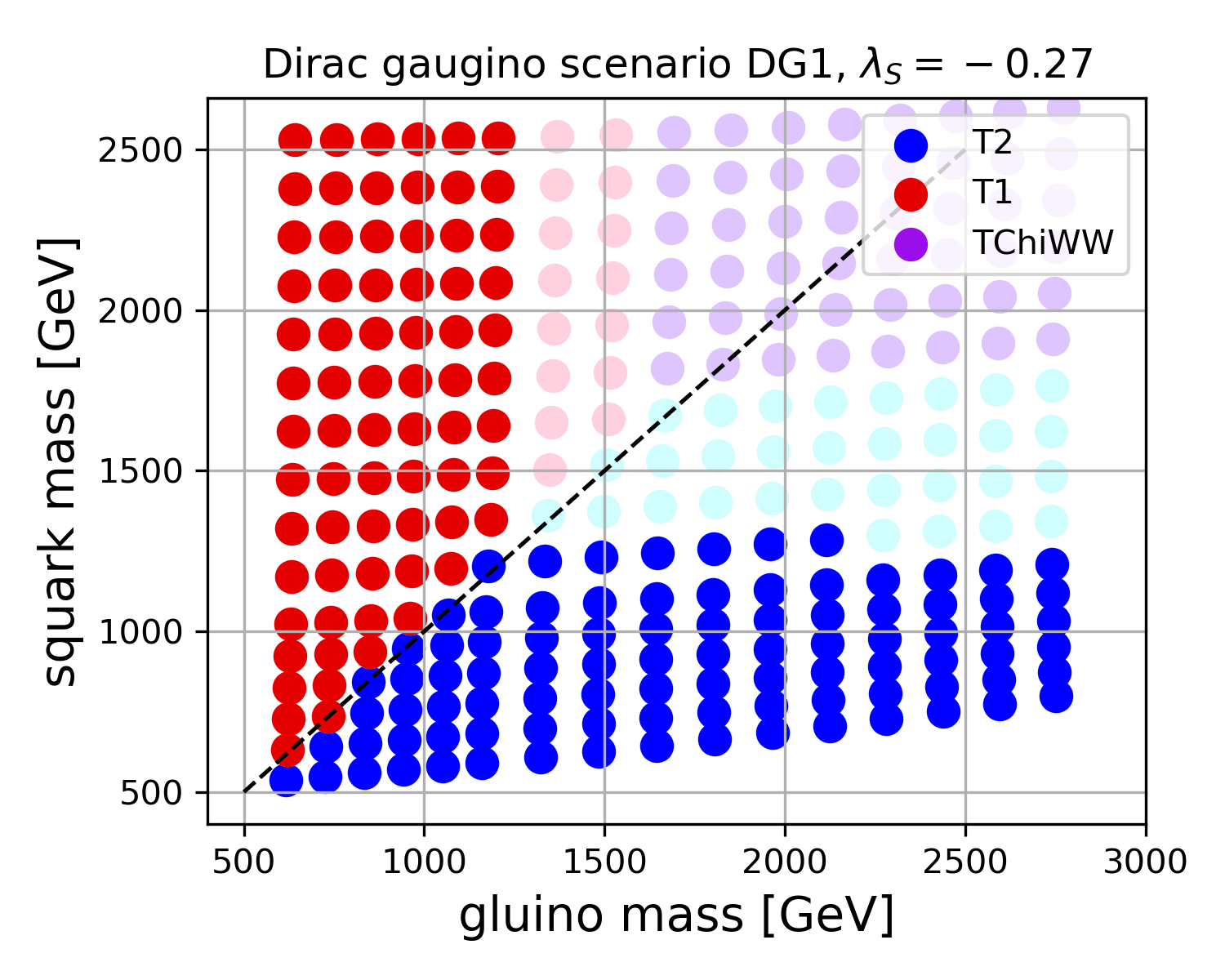}%
\includegraphics[width=0.5\textwidth]{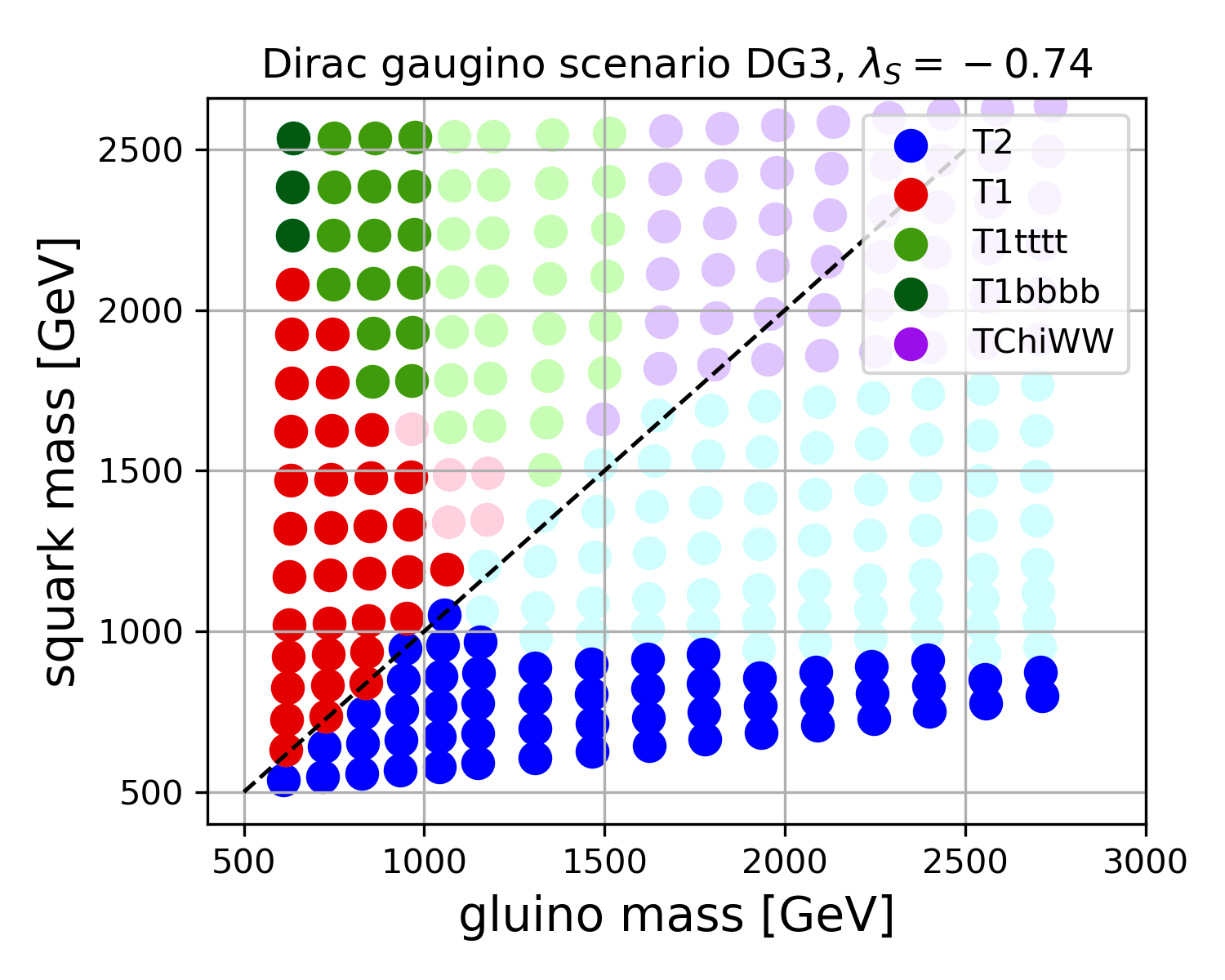}
\caption{\label{fig:smodels-exclusion}
{\tt SModelS} constraints in the gluino versus squark mass plane, on the left for DG1, on the right DG3. 
The colour code denotes the simplified model which gives the strongest constraint 
(T1: $pp\to \tilde g\tilde g$, $\tilde g \to q\bar q \tilde\chi^0_1$; T1tttt:  $pp\to \tilde g\tilde g$, $\tilde g \to t\bar t \tilde\chi^0_1$; 
T2: $pp \to \tilde q\tilde q^{(*)}$,  $\tilde q \to q \tilde\chi^0_1$; 
TChiWW: $pp \to \tilde\chi^\pm_i\tilde\chi^\pm_i$,  $\tilde\chi^\pm_i\to W^\pm \tilde\chi^0_1$).
Full-colour (non-transparent) points are excluded by {\tt SModelS}, while light-shaded points escape the simplified model limits. 
}
\end{figure}

The result is shown in Fig.~\ref{fig:smodels-exclusion}.  
For DG1, when $m_{\tilde g}<m_{\tilde q}$ the strongest constraint comes from the 
$pp\to \tilde g\tilde g$, $\tilde g \to q\bar q \tilde\chi^0_1$ simplified model (denoted as T1) and excludes gluino masses up to 
about 1250~GeV for LO cross-sections. When $m_{\tilde q}<m_{\tilde g}$, the strongest constraint  
mostly comes from the $pp \to \tilde q\tilde q^{(*)}$,  $\tilde q \to q \tilde\chi^0_1$ simplified model (denoted as T2), 
excluding squark masses up to roughly 1300~GeV as long as the gluino is not too heavy. 
In the equivalent MSSM case (MSSM1 scenario in Table~\ref{tab:branchings}), 
the gluino mass limit would be only 1 TeV due to the smaller gluino pair-production cross-section while, 
conversely, the squark mass limit would be about 2 TeV for 2.6 TeV gluinos. 

For DG3, which has stops around 1600~GeV and a $\tilde\chi^0_2$--$\tilde\chi^0_1$ mass splitting of about 35~GeV, the picture changes. 
On the one hand, over a large part of the region with $m_{\tilde g}<m_{\tilde q}$, the strongest constraint now comes from the 
$pp\to \tilde g\tilde g$, $\tilde g \to t\bar t \tilde\chi^0_1$ simplified model (denoted as T1tttt). Moreover, and more importantly, 
gluino and squark decays via the bino-like $\tilde\chi^0_{2}$ are followed by $\tilde\chi^0_{2} \to \tilde\chi^0_{1}\,f\bar f$ via an off-shell $Z$, 
which is a different topology in the simplified model picture.%
\footnote{In {\tt SModelS} txname notation, these would be constrained by, e.g., T5ZZoff or T6ZZoff results, which are however not available.} 
This drastically reduces the effective cross-section ($\sigma\times {\rm BRs}$) that goes into the T1, T1tttt or T2 topologies. % which are constrained by simplified model results. 
Consequently, the excluded region is noticeably smaller for DG3 than for DG1, 
with a gluino mass limit of only 1~TeV (corresponding to the \mbox{factor 2} reduction of the T1 cross-section which is also seen in the comparison between DG1 and MSSM1 above), and a squark mass limit below 1~TeV.

It is also worth pointing out that for heavy gluinos and squarks, the effective T1(tttt) or T2 cross-sections become too small and  
electroweak production of charginos followed by $\tilde\chi^\pm_i\to W^\pm \tilde\chi^0_1$ decays 
(denoted as TChiWW) takes over as the most constraining simplified model signature. 
Note however that TChiWW upper limit maps are available for 8~TeV only---neither ATLAS nor CMS have provided them for the 13~TeV data---and do not exclude any of the scan points.
 
%==============================================================================
\section{Recast of the ATLAS multi-jet plus $\met$ analysis}\label{sec:recast}
%==============================================================================

From the above discussion it is clear that the simplified model limits are not sufficient for constraining complex scenarios as the ones considered here. 
Instead, a full recasting of the experimental search(es) is necessary to derive the true exclusion limit. To this end, 
we have implemented the ATLAS multijet search~\cite{Aaboud:2017vwy} in {\tt MadAnalysis\,5}~\cite{Conte:2012fm,Conte:2014zja, Dumont:2014tja}. 
This is a generic search for squarks and gluinos in final states with jets and large missing transverse momentum, $\met$,  
using \mbox{36~fb$^{-1}$} of $\sqrt{s} = 13$~TeV $pp$ collision data. 
It employs two approaches: one referred to as ‘Meff-based search’ and a 
second, complementary search using the recursive jigsaw reconstruction technique. 

Here we use only the Meff-based analysis,   
which comprises 24 inclusive signal regions characterized by a minimum required jet multiplicity of 
two, four, five or six jets with transverse momenta $p_T > 50$~GeV. The missing energy of the event must be larger than 250~GeV, and 
events with a baseline electron or muon with $p_T > 7$~GeV are vetoed.  
Signal regions requiring the same jet multiplicity are distinguished by increasing background rejection through cuts in variables 
like the $p_T$ of the leading jets, $\Delta\Phi$ between jets and $\met$, and the effective mass variable $M_{\rm eff}$~\cite{Hinchliffe:1996iu} 
(defined as the scalar sum of the $p_T$ of the leading jets and the $\met$), among others. 
Of these 24 signal regions, 22 are implemented in the {\tt MadAnalysis\,5} recast code, which is publicly available as \cite{ma5:recast} and part of the {\tt MadAnalysis\,5} Public Analysis Database~\cite{Dumont:2014tja}. 
Two additional signal regions using larger-radius jets (dubbed 2jB-1600 and 2jB-2400 in the ATLAS paper) are not included as we could not reach a good enough agreement with the validation material provided by ATLAS.

To evaluate the sensitivity of this search to gluinos and squarks in the Dirac gaugino model, 
we scan over gluino and light-flavor squark masses for the four benchmark scenarios of section~\ref{sec:benchmarks}. 
For each scan point, we simulate 30K events with {\tt MadGraph5\_aMC@NLO}~\cite{Alwall:2014hca},   
including all $2\to 2$ SUSY production processes in $pp$ collisions at 13~TeV using {\tt nn23lo1} PDFs. 
Decays, parton shower and hadronization are done in {\tt Pythia\,8.2}~\cite{Sjostrand:2014zea}
and the simulation of the ATLAS detector with  {\tt Delphes\,3}~\cite{deFavereau:2013fsa}. 
Finally the events are analysed with {\tt MadAnalysis\,5} and an exclusion confidence level (CL) 
is computed with the CL$_s$ technique~\cite{Read:2002hq}. 
See \cite{Conte:2018vmg} for a comprehensive introduction to recasting with {\tt MadAnalysis\,5}, explaining the full procedure. 
Note that in each scan point only the ``best'' (i.e.\ the statistically most sensitive) 
signal region is used for limit setting.\footnote{Since the signal regions are inclusive (= overlapping) they actually cannot be combined.}

\begin{figure}[t!]\centering
   \includegraphics[width=0.78\linewidth]{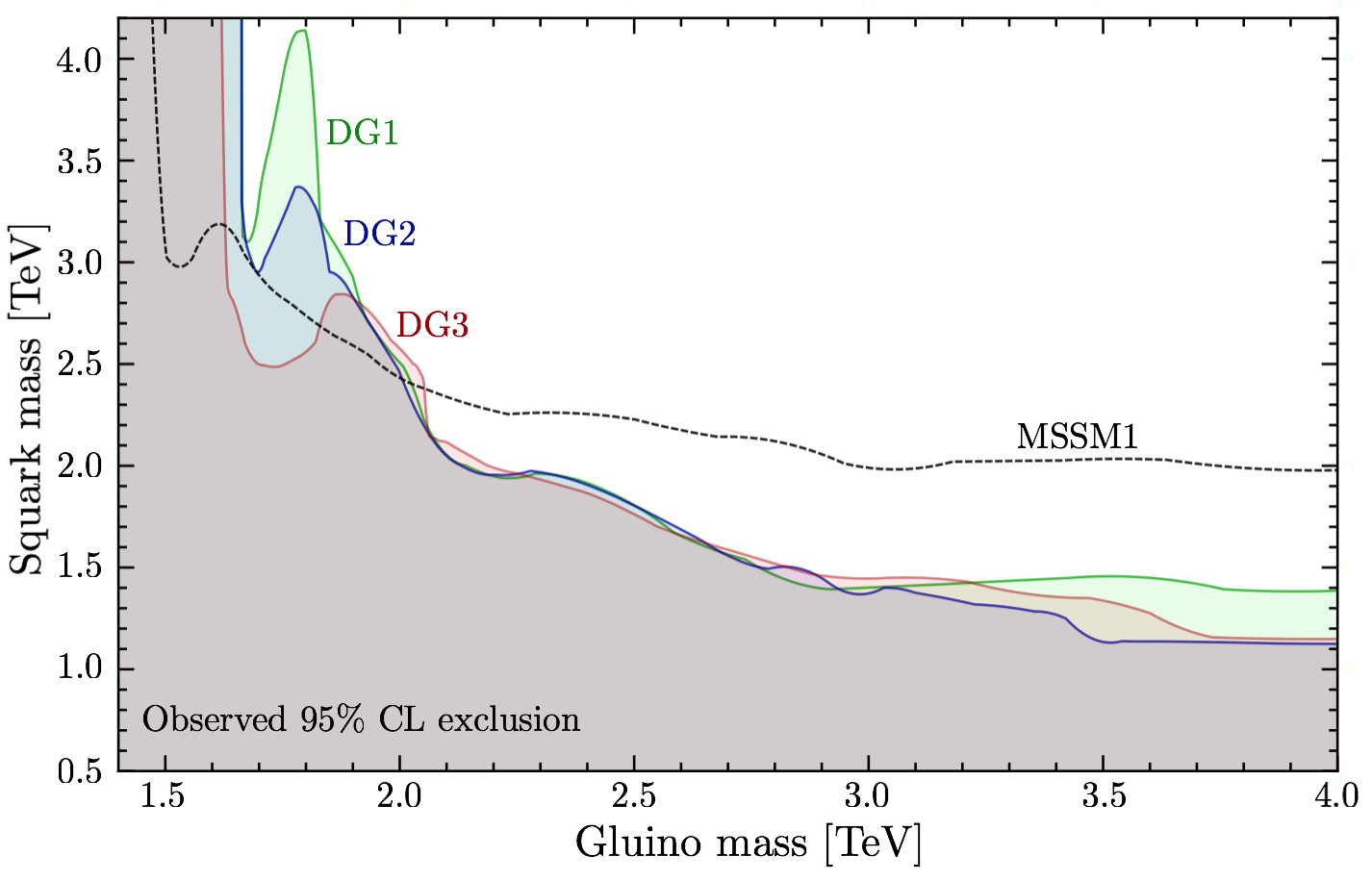}
\caption{95\% CL exclusion limits in the gluino vs.\ squark mass plane for DG1 (green), DG2 (blue) and DG3 (red) contrasted with MSSM1 (black dashed line), derived from the recasting of the ATLAS 2--6 jets + $\met$ analysis for 36~fb$^{-1}$ at $\sqrt{s}=13$~TeV. Only the most sensitive (=best expected) signal region is used for the limit setting.
}\label{fig:AllExclusions}
\end{figure}

Let us start with the light wino scenarios. 
Figure~\ref{fig:AllExclusions} shows the resulting 95\% CL exclusion lines in the gluino vs.\ squark mass plane for DG1, DG2, DG3 and MSSM1. 
As can be seen, for $m_{\tilde g}\approx m_{\tilde q}$, the limit is about 2.1~TeV for both gluino and squark masses in all DG benchmark 
scenarios. For 4~TeV gluinos, the squark mass limit is about 1.4~TeV in the least favourable DG case (DG1), decreasing to about 1.1--1.15~TeV 
for DG2 and DG3, where $\tilde\chi^0_2\to Z^*\tilde\chi^0_1$ decays appear in the squark decay chains. 
(The comparison with the MSSM will be done at the end of this section.)

\begin{figure}[t!] \centering
\begin{subfigure}[b]{0.58\textwidth}
    \caption{ DG1 \label{DG1_CLsContributions} }
   \includegraphics[width=1\linewidth]{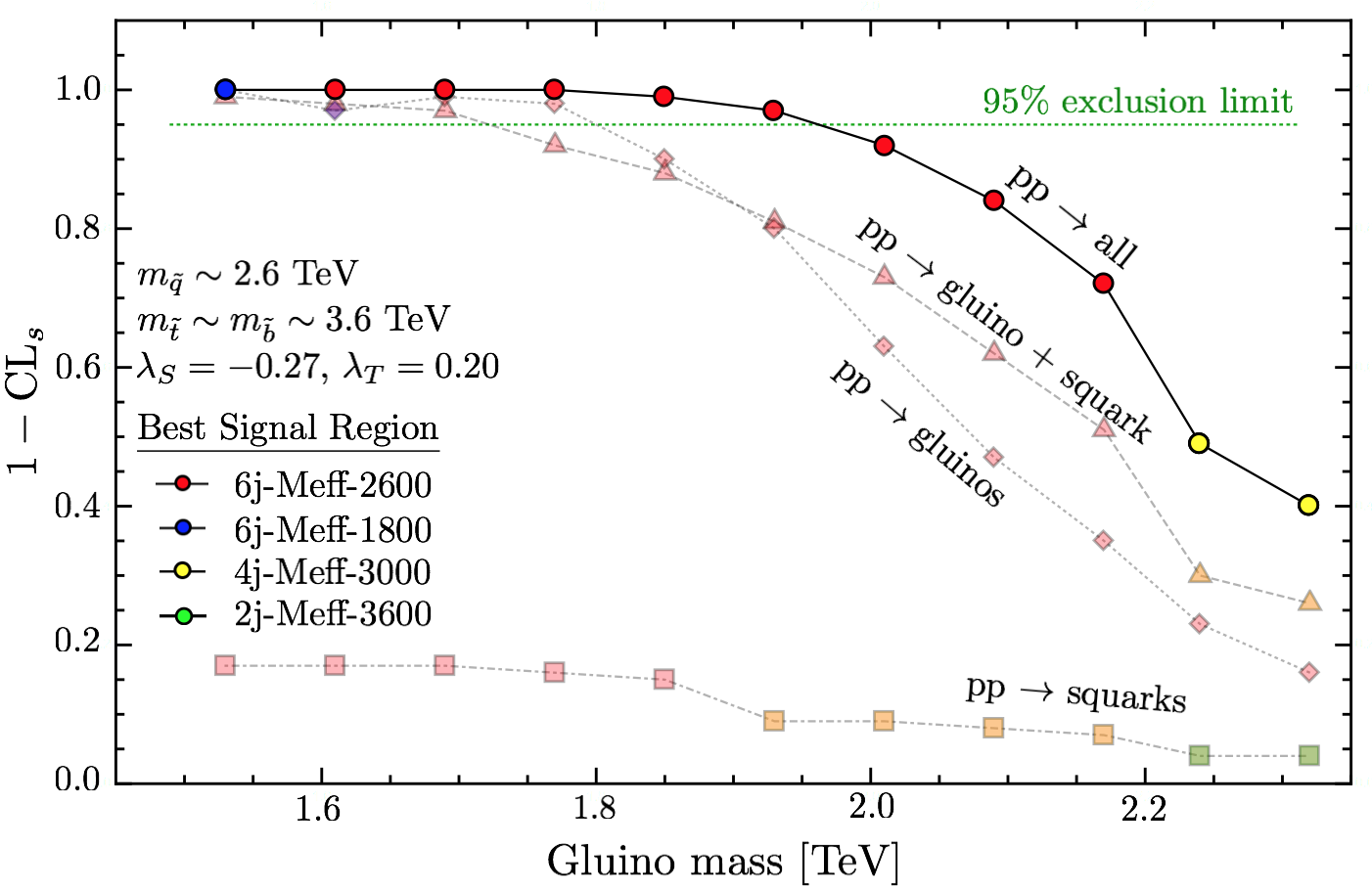}\\
\end{subfigure}
\begin{subfigure}[b]{0.58\textwidth}
      \caption{ DG2  \label{fig:DG2_CLsContributions}}
   \includegraphics[width=1\linewidth]{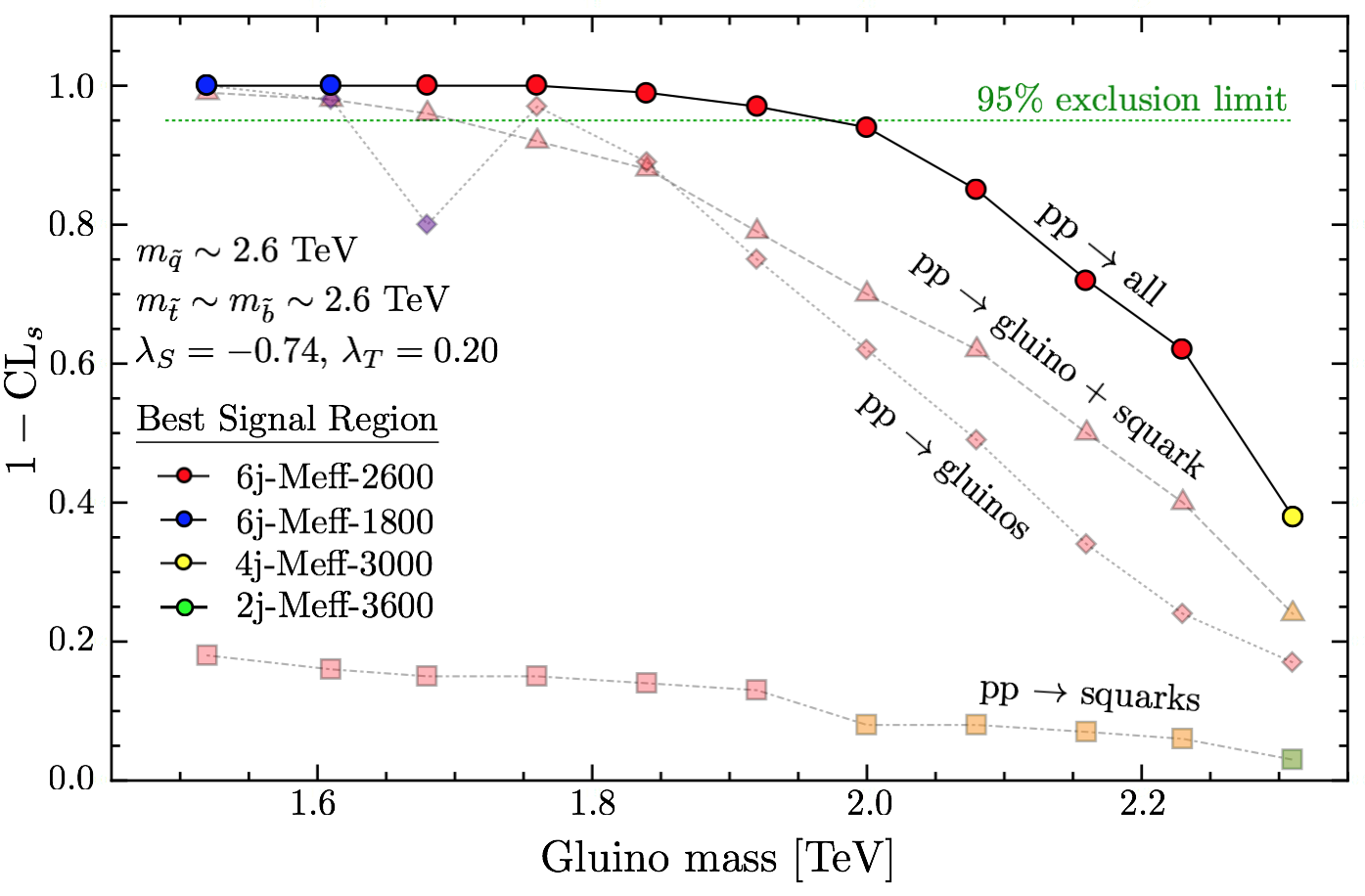}\\
\end{subfigure}
\begin{subfigure}[b]{0.58\textwidth}
   \caption{ DG3  \label{fig:DG3_CLsContributions}}
   \includegraphics[width=1\linewidth]{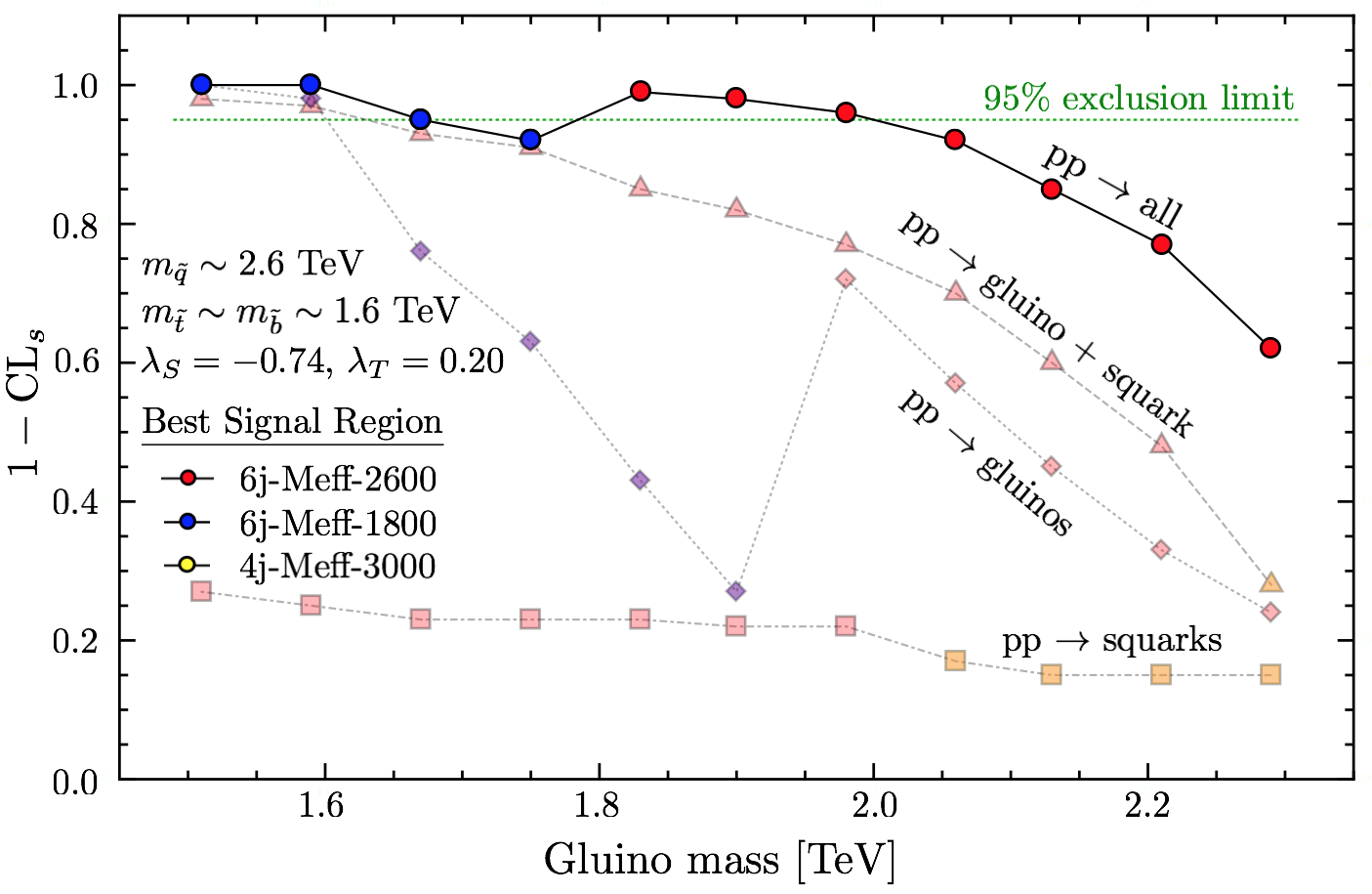}
\end{subfigure}
\caption{1-CLs values in the best signal regions from all proton-proton processes as a function of gluino mass for (a) DG1, (b) DG2, (c) DG3;  
$m_{\tilde q}\sim 2.6$~TeV in all three cases. 
Individual contributions to the total CLs (denoted by the solid black line labelled $pp \rightarrow \rm{all}$) are given by the faint dashed lines, namely gluino-pair production (diamonds); squark-pair production (triangles) and gluino-squark production (squares). 
The best signal region at each gluino mass value is identified by the colour code as indicated in the plot legends.}
\label{fig:CLs-contributions-mgluino}
\end{figure}

The gluino mass limit in the region $m_{\tilde q}>m_{\tilde g}$ depends more sensitively on the assumed DG scenario. 
While we find a robust limit of $m_{\tilde g}\gtrsim 1.65$~TeV for very heavy squarks in all cases, we also observe different ``dips'' in the exclusion 
contours for the different benchmark scenarios. 
To understand the shape of the exclusion contour, it is instructive to consider which signal regions are used for the limit setting and 
how the various production modes contribute to the final CLs value. 
To this end, Fig.~\ref{fig:CLs-contributions-mgluino} shows the CLs values in the best signal region from various proton-proton processes as a function of gluino mass, for medium heavy squarks of $m_{\tilde q}\sim 2.6$~TeV. 

We see that the best signal region switches from  6j-Meff-1800 (6~jets, $M_{\rm eff}>1800$~GeV) 
to  6j-Meff-2600 (6~jets, $M_{\rm eff}>2600$~GeV) at different values of gluino mass for the three benchmark scenarios. 
In particular for DG3 this leads to the exclusion CL dropping below 0.95 for $m_{\tilde g}\sim 1.7$~TeV, 
where gluino decays into 3rd generation squarks become dominant, and getting back above 0.95 
for $m_{\tilde g}\sim 1.8$--2~TeV.  
Moreover, we observe that taking into account gluino-pair production would only give a bound of $m_{\tilde g}\gtrsim 1.65$--$1.7$~TeV, 
as is also found in the limit of heavy squarks in Fig.~\ref{fig:AllExclusions}. 
The inclusion of both gluino-pair and gluino-squark production is essential for a correct limit setting.\footnote{This was also pointed out in \cite{Ambrogi:2017lov} in the context of simplified model limits.} 

\begin{figure}[t!] \centering
\begin{subfigure}[a]{0.64\textwidth}
   \caption{ DG1 } \includegraphics[width=1\linewidth]{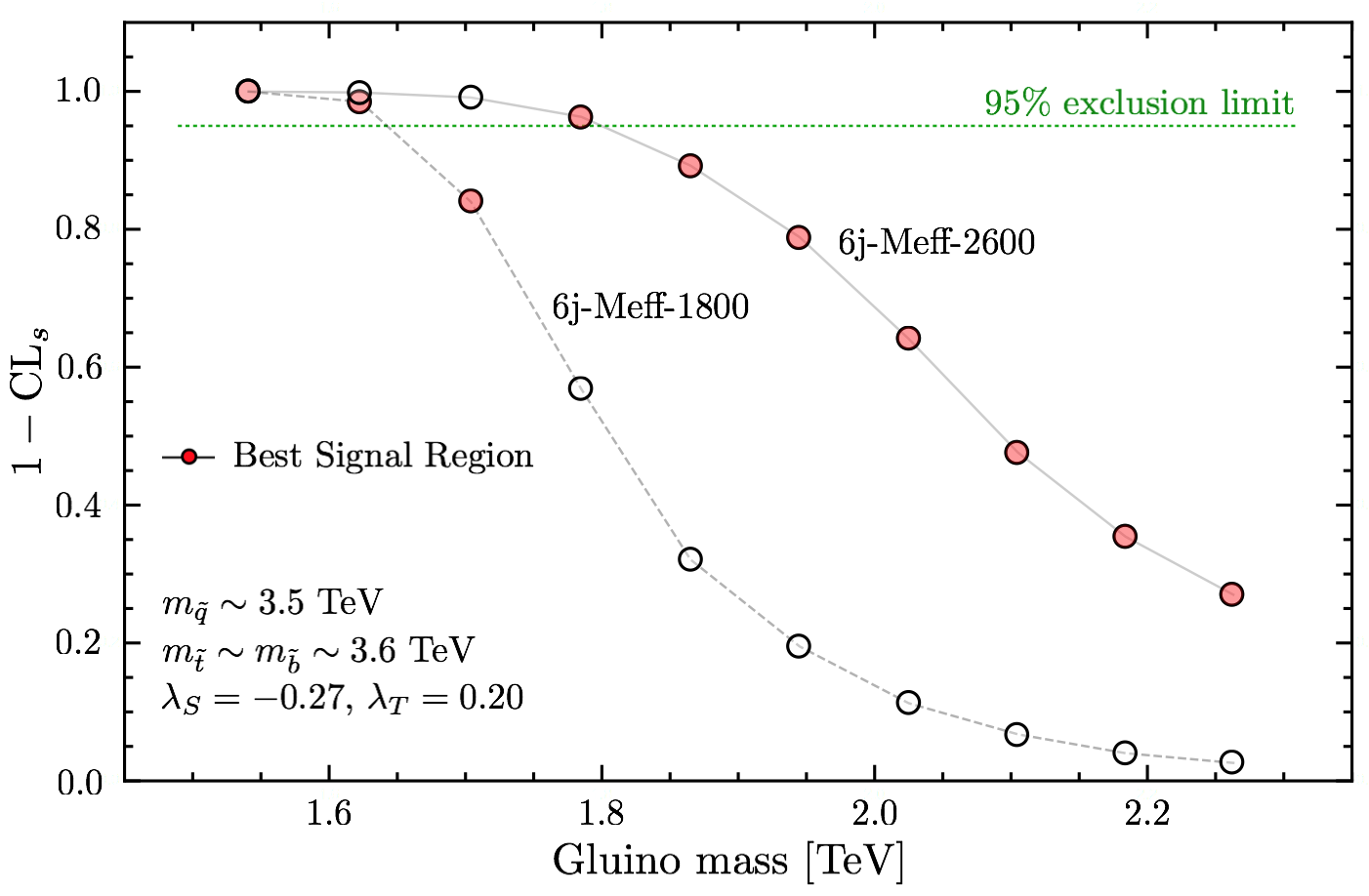}\\
\end{subfigure}
\begin{subfigure}[b]{0.64\textwidth}
   \caption{ DG3 } \includegraphics[width=1\linewidth]{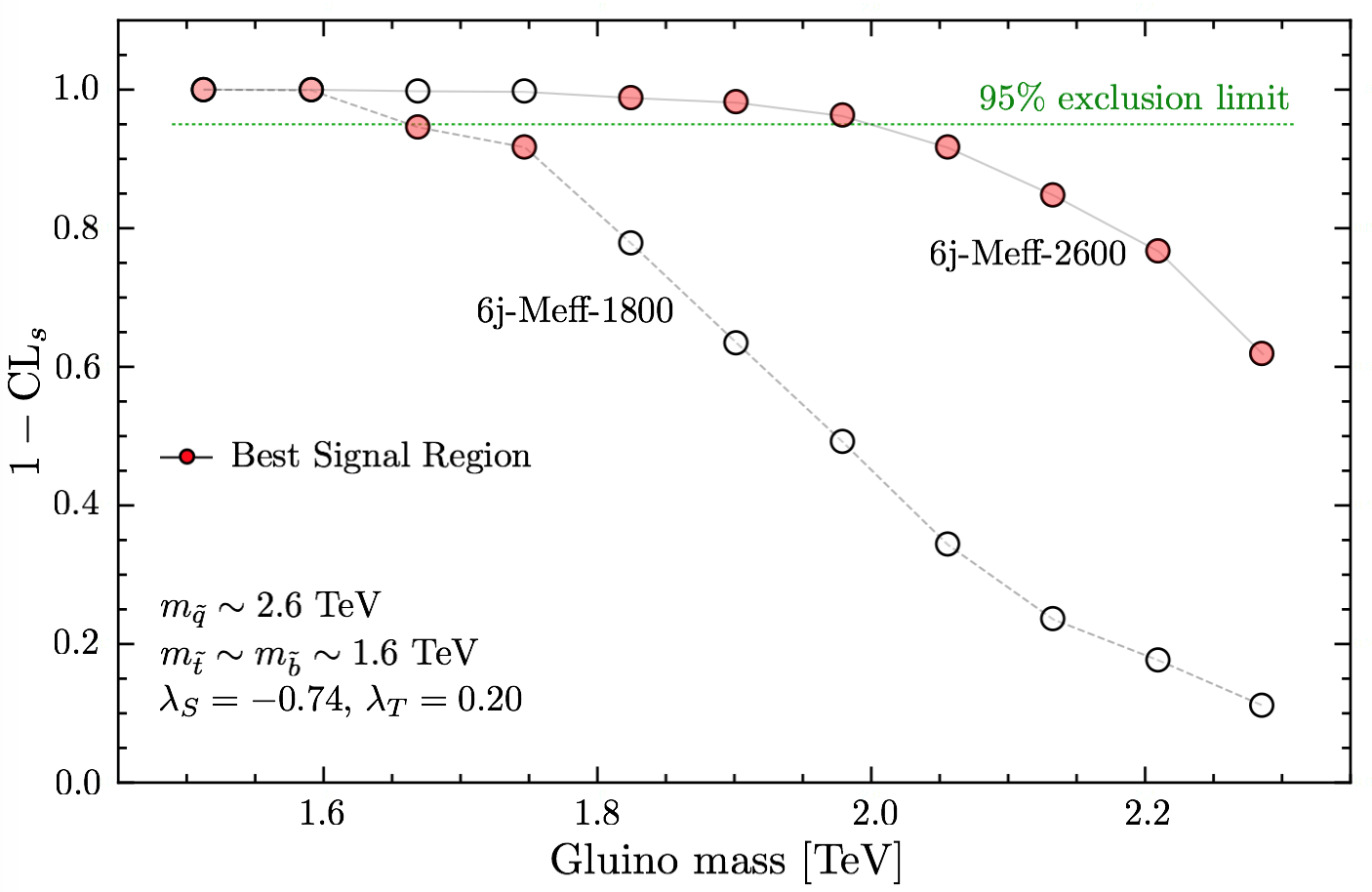}
\end{subfigure}
\caption{Comparison of 1-CLs values in the 6j-Meff-1800 and 6j-Meff-2600 signal regions as a function of gluino mass, for (a) DG1 with $m_{\tilde{q}} \sim 3.6$ TeV and (b) DG3 with $m_{\tilde q}\sim 2.6$~TeV. The best signal region is identified by full red circles.}
\label{fig:CLs-vs-mglu}
\end{figure}

Next, we compare in Fig.~\ref{fig:CLs-vs-mglu} the CLs values in different signal regions for DG1 and DG3. 
%as a function of gluino mass, for fixed squark mass. 
In order to cut across the dip-peak features in the exclusion contours, we here choose $m_{\tilde q}\sim 3.6$~TeV for DG1 and 
$m_{\tilde q}\sim 2.6$~TeV for DG3. 
We see again that for relatively light gluinos the best signal region is 6j-Meff-1800 
and the observed CL value drops below 0.95 for gluino masses around 1.65~TeV. 
The 6j-Meff-2600 signal region, on the other hand, excludes higher gluino masses, up to about 1.8 TeV in DG1 with $m_{\tilde q}\sim 3.5$~TeV, 
and up to about 2~TeV in DG3 with $m_{\tilde q}\sim 2.6$~TeV.
However, 6j-Meff-2600 becomes the ``best'' signal region (used for the limit setting in Fig.~\ref{fig:AllExclusions}) only for gluino masses of 1.8~TeV onwards. This is responsible for the dip-peak structure in the exclusion curve in Fig.~\ref{fig:AllExclusions}; 
using only the 6j-Meff-2600 signal region, the gluino mass limit would be stronger. 

\begin{figure}[t!]\centering
\begin{subfigure}[a]{0.64\textwidth}
\caption{ DG1 }   \includegraphics[width=1\linewidth]{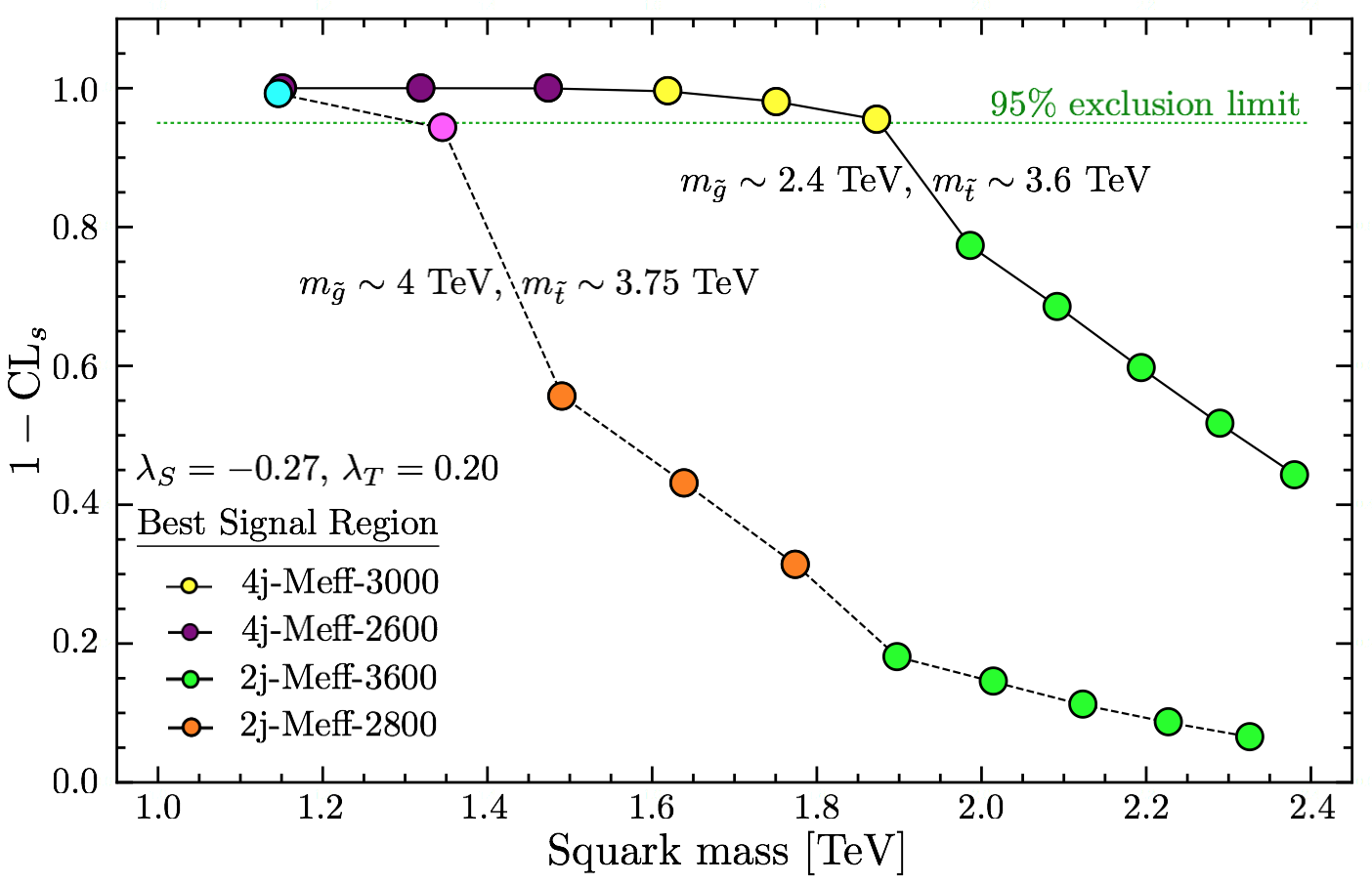} \\
\end{subfigure}
\begin{subfigure}[b]{0.64\textwidth}
   \caption{ DG3 }  \includegraphics[width=1\linewidth]{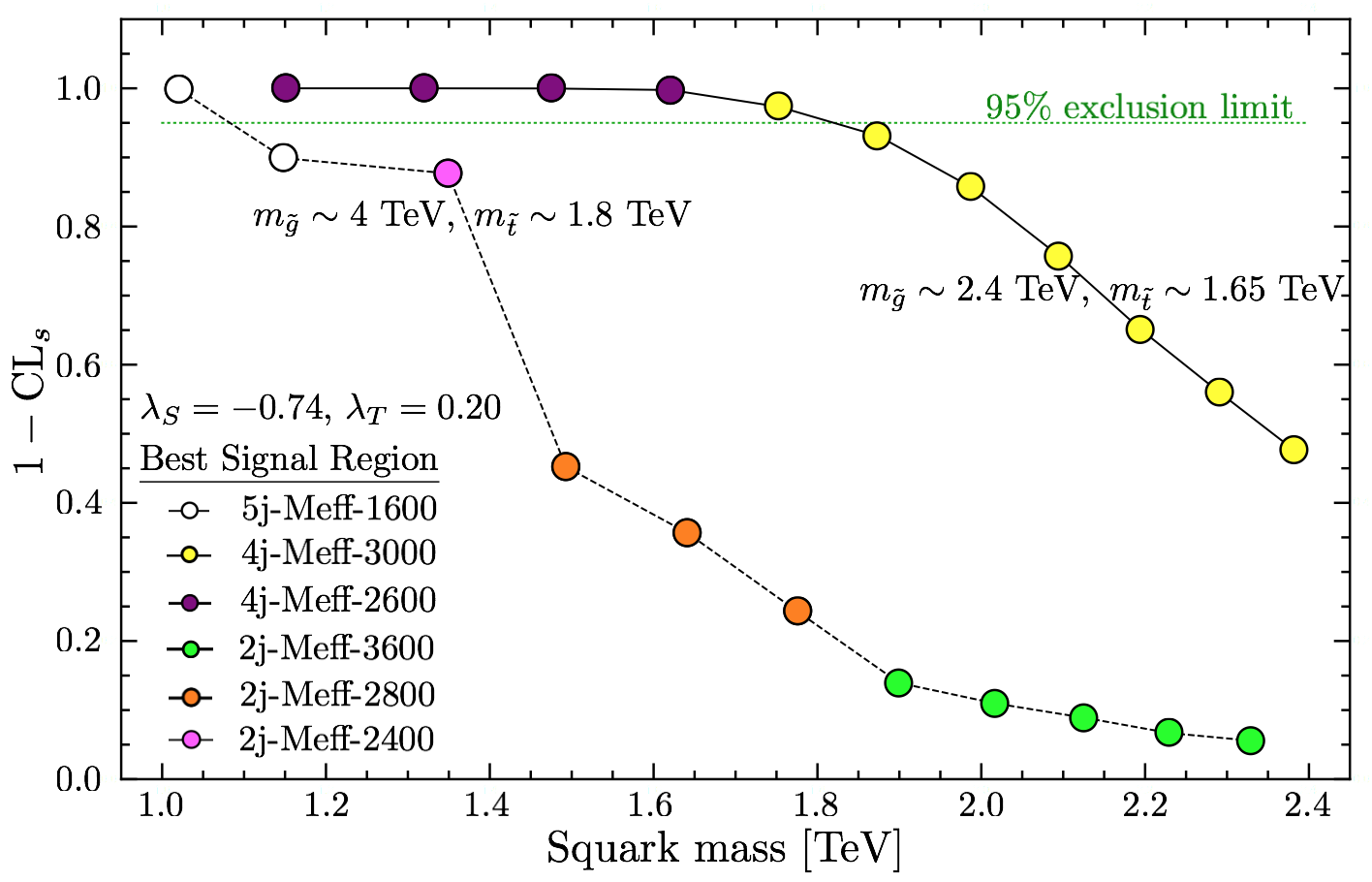}
\end{subfigure}
\caption{1-CLs values in the best signal regions from all proton-proton processes as a function of squark mass for (a) DG1 and (b) DG3. The solid lines are for $m_{\tilde{g}} \sim 2.4$ TeV, while the dashed lines are for $m_{\tilde{g}} \sim 4$ TeV. 
(Since the input parameters are the soft masses, $m_{\tilde{t}}$ and $m_{\tilde{b}}$ vary slightly in the two cases.) 
} \label{fig:CLs-contributions-msquark}
\end{figure}

Turning to the squark exclusion limits, 
Fig.~\ref{fig:CLs-contributions-msquark} shows the CLs values in the best signal regions as a function of squark mass, for fixed gluino mass. 
We again compare only DG1 and DG3, as DG2 is very similar to the latter. 
For $m_{\tilde g} \sim 2.4$~TeV, signal regions with 4~jets (first 4j-Meff-2600 and then 4j-Meff-3000) exclude squark masses up to 1.9 (1.8) TeV for DG1 (DG3). This is partly due to a substantial contribution from gluino-squark production. 
As the gluino mass is increased to $\sim$ 4~TeV, both squark-pair and gluino-squark production cross-sections are suppressed, 
and the best signal region is typically one with only 2 jets. The exception is DG3 with squark masses around 1~TeV, where a 5-jet signal region  
with rather low $M_{\rm eff}$ cut (5j-Meff-1600) becomes the best one. This is again a consequence of the  $\tilde\chi^0_2\to Z^*\tilde\chi^0_1$ decays, which are present in DG3 (and DG2) but not in DG1. 

\clearpage

\clearpage

\clearpage

Comparing all this to the equivalent MSSM1 scenario, we see the expected $\sim 200$~GeV lower gluino mass limit; the squark mass limit is however considerably stronger when gluinos are heavy, still reaching $m_{\tilde q}\gtrsim 2$~TeV for 4~TeV gluinos, as Majorana gluinos decouple very slowly. 

\begin{figure}[t!]\centering
   \includegraphics[width=0.78\linewidth]{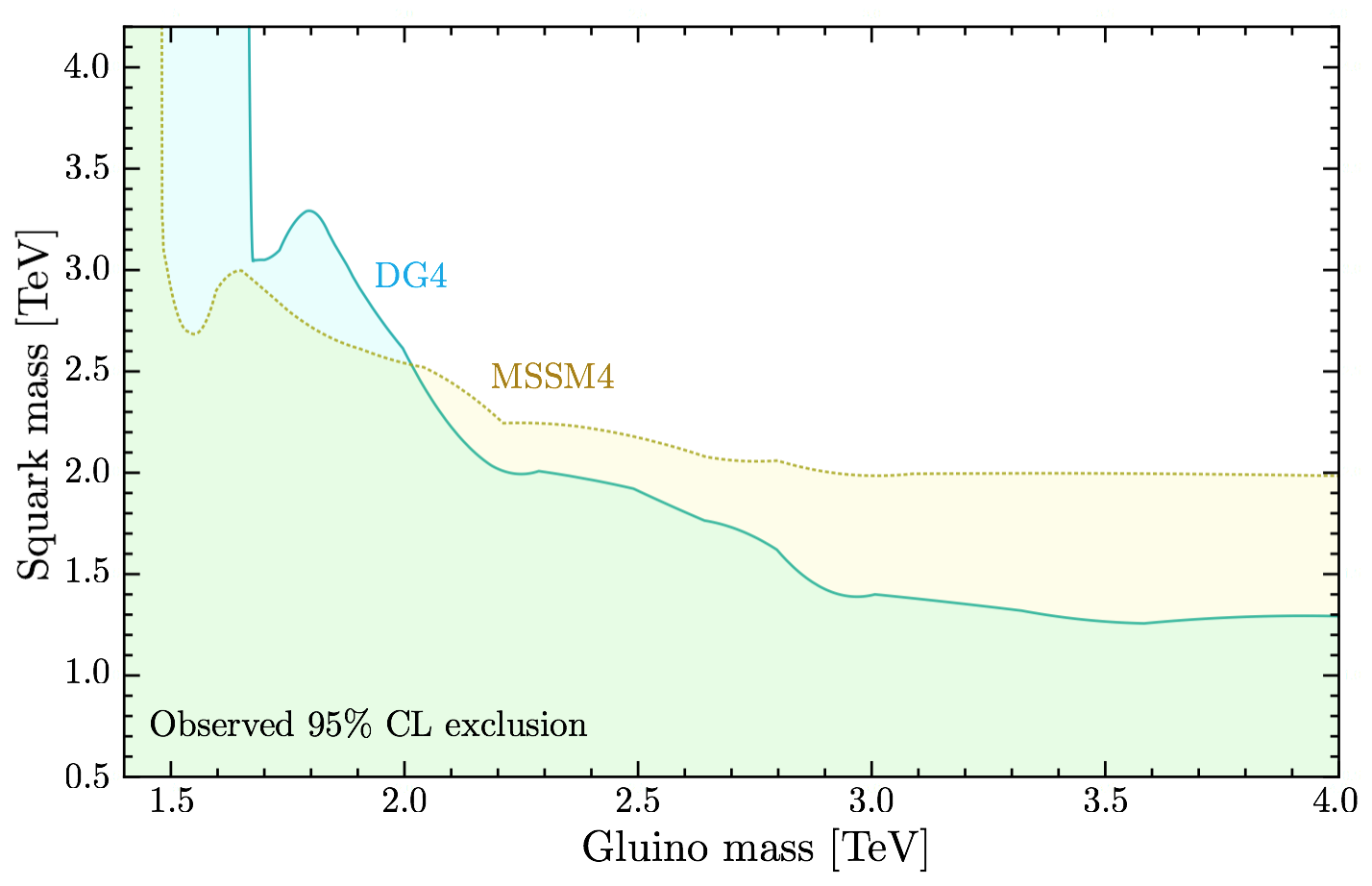}
\caption{95\% CL exclusion limits in the gluino vs.\ squark mass plane for DG4, and comparison to MSSM4.}
\label{fig:DG4Exclusions}
\end{figure}

Last but not least let us explore the role of light or heavy winos appearing in the decay chains. 
To this end, Fig.~\ref{fig:DG4Exclusions} shows the 95\% CL exclusion limits in the gluino vs.\ squark mass plane for 
the DG4 and MSSM4 scenarios, to be compared with the exclusion lines for DG2 and MSSM1 in Fig.~\ref{fig:AllExclusions}.  
Interestingly, the results are very similar for heavy and light winos; 
the main difference is an increase in the squark mass limit by about 100--200~GeV (for fixed gluino mass) when winos are heavy. 
In particular, $m_{\tilde q}\gtrsim 1.3$~TeV at $m_{\tilde g}\gtrsim 4$~TeV for DG4, which lies in between the values for DG1 and DG2,3.

\begin{figure}[t!]\centering
   \includegraphics[width=0.74\linewidth]{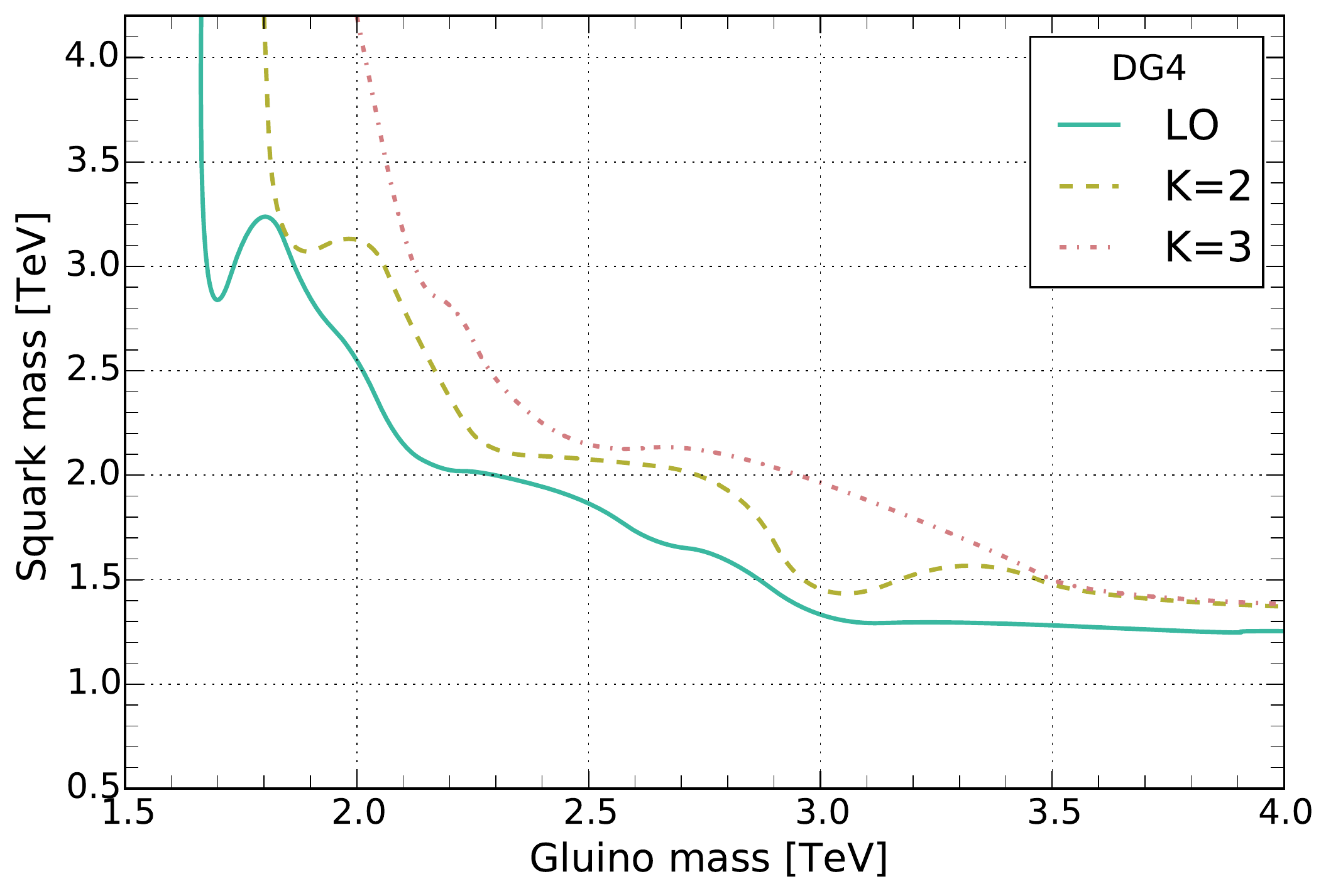}
\caption{95\% CL exclusion limits in the gluino vs.\ squark mass plane for benchmark DG4 with K-factors 1 (LO), 2 and 3.}
\label{fig:KFACTORS}
\end{figure}

Before concluding, a comment is in order on the effect of higher-order corrections. 
It is well known from the MSSM~\cite{Beenakker:1996ch,nllfast} %\cite{Beenakker:1996ch,Beenakker:1997ut,Kulesza:2008jb,Kulesza:2009kq,Beenakker:2009ha,Beenakker:2010nq,Beenakker:2011fu}, 
that K-factors for gluino-pair and gluino-squark production can be very large, 
of the order of a factor 2--3, depending on the PDF set used; K-factors for squark production are somewhat smaller but still sizeable. 
For the DG case, the next-to-leading order (NLO) corrections to squark production in the R-symmetric model 
were computed in \cite{Diessner:2017ske},  
with the conclusion that NLO K-factors are generally larger than in the MSSM by the order of 10--20\%. 
Since the cross-section of squark production falls off very steeply with increasing squark mass, $\rm{K}\approx2$   
has only little impact, pushing the gluino mass limit about 100~GeV higher. 
The higher-order corrections for Dirac gluino final states have not been computed explicitly, but we may assume they 
are not vastly different from the MSSM. Taking a K-factor of 2--3 as the reference, the gluino mass limit increases 
by roughly 200~GeV to $m_{\tilde g}\gtrsim 2$~TeV for heavy squarks, while for $m_{\tilde g}\approx m_{\tilde q}$ 
the limit is pushed to roughly 2.3--2.4~TeV. 
We illustrate this explicitly for the scenario DG4 in Fig.~\ref{fig:KFACTORS}. 
 
%==============================================================================
\section{Conclusions}\label{sec:conclusions}
%==============================================================================

Most SUSY searches at the LHC are optimised for the MSSM, where gauginos are Majorana particles. 
Dirac gauginos are, however, an interesting and theoretically well-motivated alternative. 
Their phenomenological consequences at the LHC include that gluino-pair production is enhanced 
by a factor 2 as compared to the MSSM, while squark production is strongly suppressed due to a much 
faster decoupling of the gluino t-channel exchange. 
Moreover, the extended chargino and neutralino sector present in DG models can have important effects 
on the collider signatures. 
  
In this paper, we have investigated the bounds from LHC searches on squarks and gluinos in the Minimal Dirac Gaugino 
Supersymmetric Standard Model for several representative benchmark scenarios. 
Since a typical MDGSSM scenario should have electroweakinos not too far above the electroweak scale, we chose, 
as a primary test case, scenarios with a bino-like LSP around 200~GeV, higgsinos around 400~GeV and winos around 500~GeV.  
Thus all charginos and neutralinos may appear in gluino and squark cascade decays.  We also considered a scenario 
with heavier winos of about 1200~GeV, and we compared all these to the nearest equivalent models in the MSSM.

In the context of simplified model constraints, derived with {\tt SModelS}, the large variety of possible decay modes 
in our benchmark scenarios led to very weak limits. The reason is, that in complex scenarios like the ones considered here, 
only a small fraction of the total SUSY production leads to simple signal topologies which are constrained by the available 
simplified model results. 

We therefore went on to confront our benchmark scenarios with a full recasting of the ATLAS multi-jet $\met$ 
search~\cite{Aaboud:2017vwy} with {\tt MadAnalysis\,5}.   
By comparing the bounds in the DG benchmark scenarios to those in the MSSM, we confirmed and quantified by how much supersoft models are supersafe: for large gluino masses, the bounds on squarks are very significantly (by several hundred GeV) suppressed compared to the MSSM, and this should have consequences for the naturalness of allowed models. We showed that this statement is robust even including loop corrections to the production. On the other hand, for smaller gluino masses, the extra degrees of freedom lead to larger production cross-sections, and so the lower limit on the gluino mass in these models is somewhat higher than in the MSSM. 

An important feature of the DG case, which we discussed in some detail in this paper, is that the trilinear $\lambda_S$ and $\lambda_T$ couplings, 
which give a tree-level boost to the light Higgs mass, lead to small mass splittings within the bino and wino states. 
This is important for LHC phenomenology because, if the mass splitting between the two lightest states (in our benchmark scenarios 
the two binos) is very small, then the $\tilde\chi^0_2$ can live long enough to effectively be a co-LSP on collider scales and 
appear only as $\met$.
For larger mass splittings, however, the $\tilde\chi^0_2$ may decay promptly into $f\bar f\tilde\chi^0_1$ via an off-shell $Z$-boson, 
leading to an additional step in part of the gluino and squark cascade decays.  
For $m_{\tilde g}\approx m_{\tilde q}$ this has no noticeable influence on the mass limits. For heavy gluinos or squarks, however, 
we showed that the mass limits slightly weaken when $\lambda_S$ is large. Last but not least, there exists a range of $\lambda_S$ where the $\tilde\chi^0_2$ is a long-lived neutral particle, 
whose decays can give signatures with displaced vertices. A detailed study of this case is left for future work. 
 
\bigskip

\noindent
{\it For the sake of reproducibility of our study, we provide ample material on Zenodo~\cite{zenodo:bm-dataset,zenodo:scanpoints}.}

%==============================================================================
\acknowledgments
%==============================================================================

This work was supported in part by the IN2P3 project ``Th\'eorie -- LHCiTools''. 
This work has also been done within the Labex ILP (reference ANR-10-LABX-63) part of the Idex SUPER, and received financial state aid managed by the Agence Nationale de la Recherche, as part of the programme Investissements d'avenir under the reference ANR-11-IDEX-0004-02, and the Labex ``Institut Lagrange de Paris'' (ANR-11-IDEX-0004-02,  ANR-10-LABX-63) which in particular funds the scholarship of SLW. 
MDG acknowledges the support of the Agence Nationale de Recherche grant ANR-15-CE31-0002 ``HiggsAutomator.''
HRG is funded by the Consejo Nacional de Ciencia y Tecnología, CONACyT, scholarship no.\ 291169.

%\clearpage
%==============================================================================
\bibliographystyle{utphys}
\bibliography{references}
%==============================================================================
%\begin{thebibliography}{99}
%\end{thebibliography}

\end{document}